\documentclass[%
reprint,
 amsmath,amssymb,
 aps,
 prl,
]{revtex4-1}

\usepackage{amsmath,amsbsy}    % need for subequations
\usepackage{graphicx}   % need for figures
\usepackage{verbatim}   % useful for program listings
\usepackage{color}      % use if color is used in text
\usepackage{subfigure}  % use for side-by-side figures
\usepackage{tabularx}
 
\def\lowsim{\mathrel{\lower 0.7 ex \hbox to 0 pt{$\sim$\hss}}}

 \newcommand{\pid}{-\Pi_{ij}D_{ij}}

 \newcommand{\m}[1]{\mathbf #1} % m stands for mota -> hindi for fat
 \newcommand{\jde}{\m{j}\cdot \m{E}}

\begin{document}
\title{Statistics of Kinetic Dissipation in Earth's Magnetosheath - MMS Observations}

\author{Riddhi Bandyopadhyay}
\affiliation{Department of Physics and Astronomy, University of Delaware, Newark, Delaware 19716, USA}
%\affiliation{Bartol Research Institute, University of Delaware, Newark, Delaware 19716, USA}	

\author{William~H. Matthaeus}
\affiliation{Department of Physics and Astronomy, University of Delaware, Newark, Delaware 19716, USA}
\affiliation{Bartol Research Institute, University of Delaware, Newark, Delaware 19716, USA}	

\author{Tulasi~N. Parashar}
\affiliation{Department of Physics and Astronomy, University of Delaware, Newark, Delaware 19716, USA}
\affiliation{Now at: School of Chemical and Physical Sciences, Victoria University of Wellington, Kelburn, Wellington 6012, NZ}

\author{Yan Yang}
\affiliation{\affiliation{3}{Southern University of Science and Technology, Shenzhen, Guangdong 518055, China}}

\author{Alexandros Chasapis}
%%\affiliation{Department of Physics and Astronomy, University of Delaware, Newark, Delaware 19716, USA}
\affiliation{Laboratory for Atmospheric and Space Physics, University of Colorado Boulder,  Boulder, Colorado, USA}	

\author{Barbara~L. Giles}
\affiliation{NASA Goddard Space Flight Center, Greenbelt, Maryland 20771, USA}

\author{Daniel~J. Gershman}
\affiliation{NASA Goddard Space Flight Center, Greenbelt, Maryland 20771, USA}

\author{Craig~J. Pollock}
\affiliation{Denali Scientific, Fairbanks, Alaska 99709, USA}

\author{Christopher~T. Russell}
\affiliation{University of California, Los Angeles, California 90095-1567, USA}

\author{Robert~J. Strangeway}
\affiliation{University of California, Los Angeles, California 90095-1567, USA}

\author{Roy~B. Torbert}
\affiliation{University of New Hampshire, Durham, New Hampshire 03824, USA}

\author{Thomas~E. Moore} 
\affiliation{NASA Goddard Space Flight Center, Greenbelt, Maryland, USA}
	
\author{James~L. Burch}
\affiliation{Southwest Research Institute, San Antonio, Texas 78238-5166, USA}

\begin{abstract}
A familiar problem in space and astrophysical plasmas
is to understand how dissipation and heating occurs. These effects are often attributed to the cascade of broadband turbulence which transports energy from large scale reservoirs 
to small scale kinetic degrees of freedom.  
When collisions are infrequent, local thermodynamic equilibrium is not established.
In this case the final stage of
energy conversion becomes more complex 
than in the fluid case, 
and both pressure-dilatation and 
pressure strain interactions (Pi-D $\equiv \pid$) 
become relevant and potentially 
important. 
Pi-D in plasma turbulence has been studied so far primarily using simulations. The present 
study provides 
a statistical analysis of Pi-D in the Earth's magnetosheath
using the unique measurement capabilities of the Magnetospheric Multiscale (MMS) mission. 
We find that the statistics of Pi-D in this naturally occurring plasma environment exhibit strong resemblance to previously established fully kinetic simulations results. The conversion of energy is concentrated in space and occurs {\em near} intense current sheets, but 
{\em not within} them. This supports recent suggestions that the chain of energy transfer channels involves regional, rather than pointwise, correlations. 
\end{abstract}

\maketitle

%\section{Introduction}
The study of dissipation processes in space and astrophysical plasmas 
is of great 
significance both 
as a fundamental plasma physics problem and 
due to its implications for
observed macroscopic effects.
In the case of weakly-collisional 
dynamics, 
typical of 
these 
space and astrophysical plasmas 
\cite{QuataertAN03, Marsch06},
fluid closures become questionable 
and fully kinetic treatment 
is required~\cite{QuataertAN03, WangJGR01}.
For weak collsionality,
the usual sign-definite dissipation functions
that emerge from Chapman-Enskog ordering 
are no longer applicable and consequently, the 
entire subject of dissipation of turbulence
and subsequent heating 
becomes challenging and even elusive. 
\begin{figure}[ht!]
    \centering
    \includegraphics[width=\linewidth]{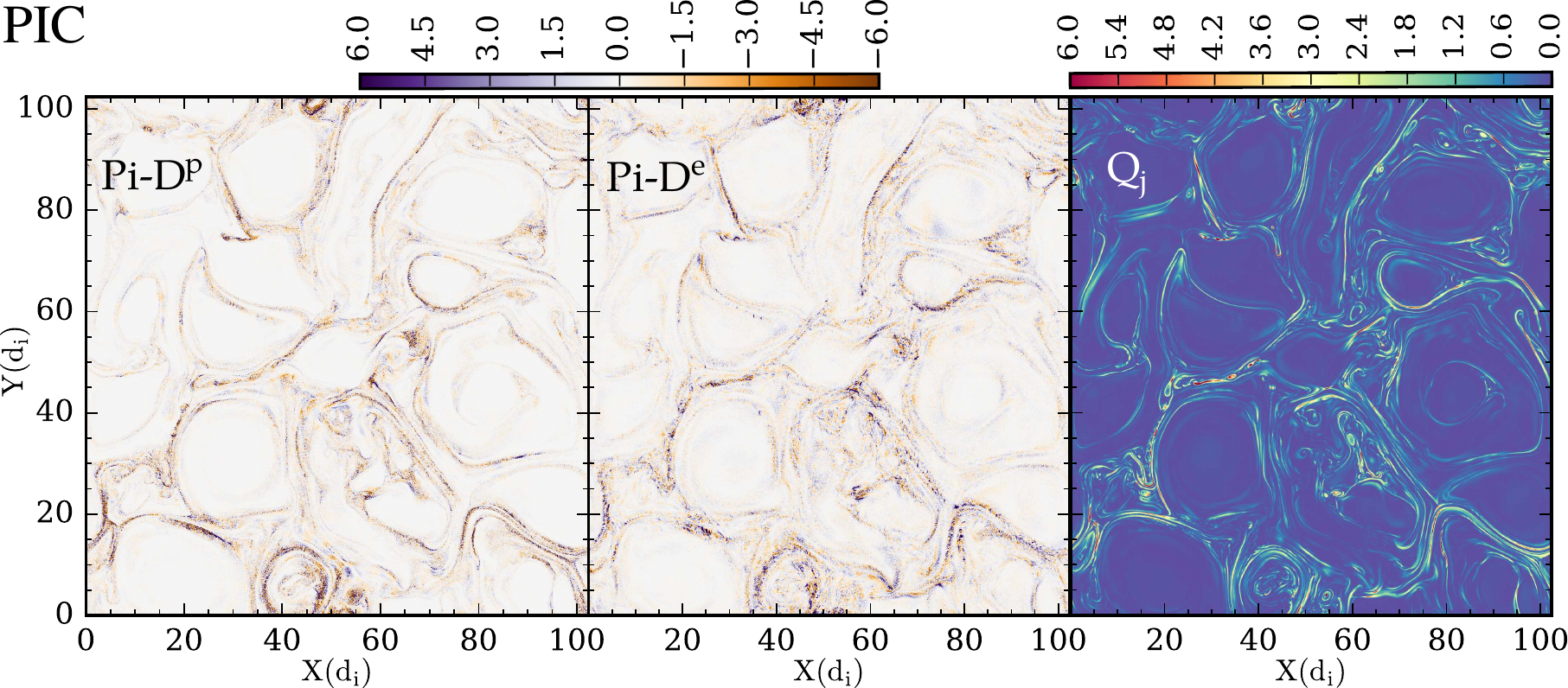}
    \includegraphics[width=\linewidth]{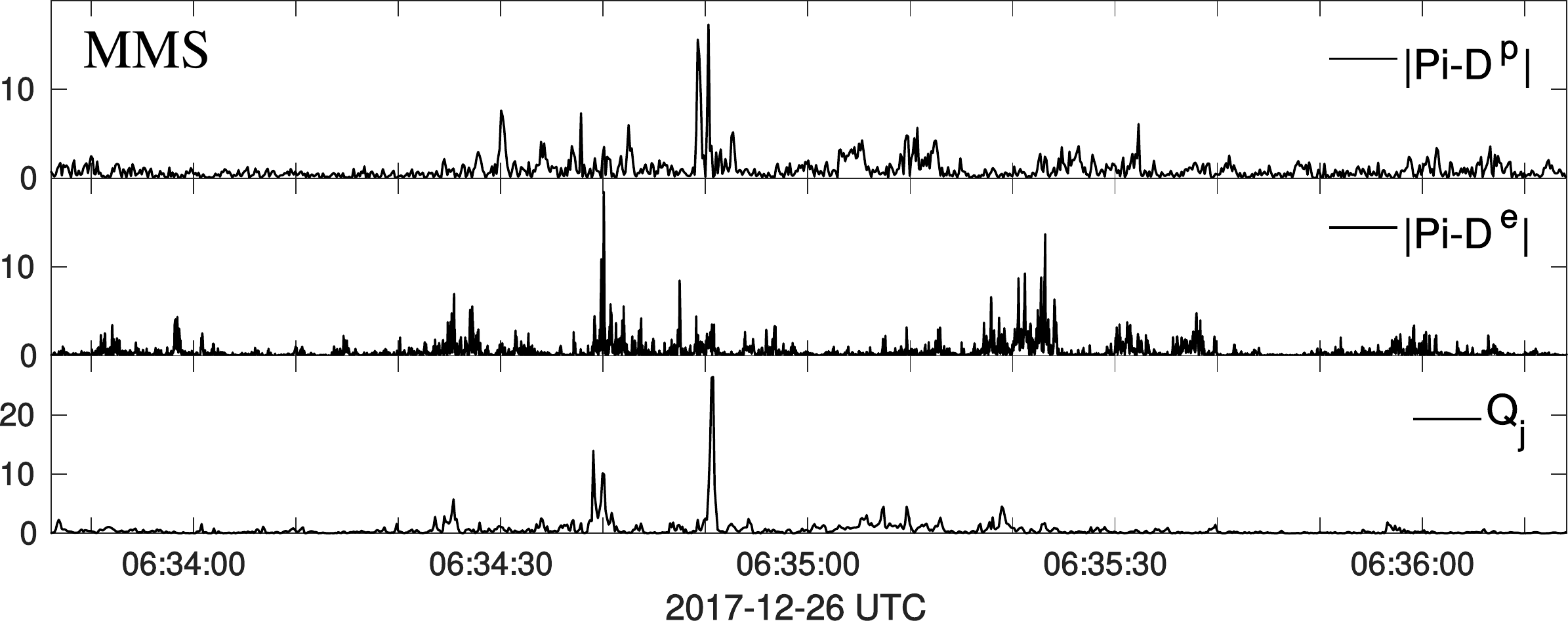}
    \caption{Normalized Pi-D, $- \Pi_{ij} D_{ij} / (- \Pi_{ij} D_{ij})_{\mathrm{rms}}$, for proton and electron, and normalized current, $Q_{\mathrm{j}} = (1/4){\mathbf{j}}^2/\langle {\mathbf{j}}^2 \rangle$ from PIC simulations (top) and a sample of MMS data (bottom).}
    \label{fig:timeseries}
\end{figure}
Even if turbulent dissipation is considered a leading candidate for 
explaining the heating of space plasmas, 
questions remain, such as:
{\it What are the rates of transfer of energy through
the available kinetic channels?}, or perhaps, 
{\em How is the turbulent fluctuation energy transferred into internal degrees of freedom of various plasma species?} 
We examine these questions, adopting a statistical approach, 
using the unique capabilities
of the 
Magnetospheric Multiscale (MMS) mission~\citep{Burch2016SSR, Pollock2016SSR, Russell2016SSR}. 
We are particularly interested in comparing the observational results with recently reported similar analyses obtained 
from kinetic plasma  simulation~\citep{Yang2017PRE,Yang2017PoP}, and this 
direct approach is enabled by the 
high-resolution, multi-spacecraft data that MMS provides.

When equations of energy exchange are computed from the hierarchy of the Vlasov-Maxwell equations, 
one finds~\cite{Yang2017PRE, Yang2017PoP}, for each species, 
here labelled by $\alpha$, 
\begin{eqnarray}
\partial_t \mathcal{E}^{f}_\alpha + \nabla \cdot \left( \mathcal{E}^{f}_\alpha \mathbf{u}_\alpha + \mathbf{P}_\alpha \cdot \mathbf{u}_\alpha \right) &=& \left( \mathbf{P}_\alpha \cdot \nabla \right) \cdot \mathbf{u}_\alpha+ \nonumber \\
& & n_\alpha q_\alpha \mathbf{E} \cdot \mathbf{u}_\alpha. \label{Eq.fenergy} \\
\partial_t \mathcal{E}^{th}_\alpha + \nabla \cdot \left( \mathcal{E}^{th}_\alpha \mathbf{u}_\alpha + \mathbf{h}_\alpha \right) &=& -\left( \mathbf{P}_\alpha \cdot \nabla \right) \cdot \mathbf{u}_\alpha. \label{Eq.thenergy} \\
\partial_t \mathcal{E}^{m} + {\frac{c}{4\pi}} \nabla \cdot \left( \mathbf{E} \times \mathbf{B} \right) &=& -\mathbf{E} \cdot \mathbf{j} \label{Eq.benergy}
\end{eqnarray}
where $q_\alpha$ is the charge, $n_\alpha$ is the number density, $\mathbf{u}_\alpha$ is the velocity, $\mathcal{E}^f_\alpha$ is the flow energy, $\mathbf{P}_\alpha$ is the pressure tensor, $\mathcal{E}^{th}_\alpha$ is the trace of pressure tensor designating internal energy, and $\mathbf{h}_\alpha$ is the heat flux for the species $\alpha$. $\mathcal{E}^m$ is the electromagnetic energy, $\mathbf{E}$ is the electric field, $\mathbf{B}$ is the magnetic field, and $\m{j}$ is the current density. The divergence terms are responsible for transporting energy spatially 
but they do not convert energy from one form to another. Furthermore, their effects integrate (by Gauss's law) to a surface effect for any finite volume. Therefore
they have 
no net contribution
for infinite (or very large) 
system size or for periodic boundary conditions (relevant for simulations).

{The basic physics embodied in Eqs.~(\ref{Eq.fenergy}-\ref{Eq.benergy}) is as follows: The term that converts energy between EM fields and particles is the well known $\jde$ term. However it is clear from Eqs.~\ref{Eq.fenergy} \& \ref{Eq.thenergy} that $\jde$ only converts energy between fields and the bulk flow of each species of particles, but not into the internal energy. The only term that  converts energy into internal energy is the pressure strain interaction PS $=-\left( \mathbf{P}_\alpha \cdot \nabla \right) \cdot \mathbf{u}_\alpha$ that converts bulk flow energy into internal energy of each species. This \textit{conversion of form of energy} into internal energy is what we mean by ``dissipation.'' This effect has been shown~\citep{Yang2017PoP, YangMNRAS18} to occur at kinetic scales, hence the terminology ``kinetic dissipation."}

The PS interaction can be further decomposed into two parts: $-\left ({\mathbf P} \cdot \nabla \right )  \cdot {\mathbf u} = -p\delta_{ij} \partial_j u_i - (P_{ij} -p\delta_{ij})\partial_j u_i = - p \theta - \Pi_{ij}D_{ij};$ where $p=\frac13 P_{ii}$, $\Pi_{ij} = P_{ij} - p\delta_{ij}$, $\theta=\nabla \cdot {\mathbf u}$ and $D_{ij} = \frac12 \left ( \partial _i u_j + \partial _j u_i \right ) - \frac13 \theta \delta_{ij}$. {Here, $\delta_{ij}$ is the Kronecker delta function.} The $p\theta$ term is the familiar dilatation term responsible for compressive heating/cooling in fluid models. The term involving the traceless tensor $\mathbf \Pi$ becomes the viscous term via the Chapman-Enskog expansion in the collisional limit. In case of collisionless systems, this term does not have a closure but can be explicitly evaluated in simulations and multi-spacecraft data sets such as MMS. We call this $\pid$ term, including the $``-"$ sign, as the ``Pi-D'' interaction~\cite{DelSartoPRE16, DelSartoMNRAS17}.

Pi-D acts intermittently in kinetic plasmas near intense intermittent structures such as strong current sheets, reconnection sites \cite{Yang2017PoP,
SitnovGRL18, YangMNRAS18, PezziApJ19}, 
{
and vorticity concentrations \cite{DelSartoPegoraro16}.} 
Shearing magnetic islands produce intense current sheets, which in turn produce quadrupole vortex structures nearby \cite{MatthaeusGRL82, Parashar2016ApJ}. Vorticity is the antisymmetric part of the velocity strain tensor and does not
contribute to a 
full contraction
 with the symmetric tensor $\Pi_{ij}$. {In plasma turbulence, vorticity concentrations can be produced by velocity shear, as it occurs in hydrodynamics, and also by reconnection-like activity  near current sheets, which is known to produce nearby quadrupolar vortex structures in both 2D~\citep{MatthaeusGRL82, Parashar2016ApJ} and 3D~\citep{Zhdankin2013ApJ, Wan2014ApJ} numerical experiments. In large Reynolds number turbulence these vortices are stretched into sheet-like structures, generating symmetric strain $D_{ij}$ \cite{Parashar2016ApJ, Yang2017PRE, DelSartoMNRAS17}. The association \cite{Parashar2016ApJ} of vortex structures, co-located concentrations of symmetric strain, and nearby electric current density has been demonstrated in 2D and 3D simulations~\cite{Parashar2016ApJ,Yang2017PoP,ChasapisApJ18}.
This complex set of dynamical couplings appears to be 
generic, and provides an explanation for the 
connection between vorticity and heating~\cite{FranciSW14, Parashar2016ApJ, DelSartoMNRAS17}. {Notably, recent magnetosheath observations have revealed a new type of coherent structures, namely electron vortex magnetic holes~\citep{Huang2017bApJL,Huang2017JGR,Huang2018ApJ}, which show correlation of electron vorticity with the increase of electron temperature, making them a possible candidate for electron heating.}} 

%\section{Data \& Results}
{To cover a large statistical sample of the turbulent magnetosheath plasma, here we focus on a 40-minute MMS burst-mode interval between 06:12:43 and 06:52:23 UTC on 26 December 2017, encompassing several $(\sim 400)$ correlation scales. At this time, the interplanetary solar wind had an average magnetic field of $ 6\,\mathrm{nT}$, flow speed $ 450\,\mathrm{km\, s^{-1}}$, and density $ 6$ cm$^{-3}$. {The MMS spacecraft, separated by $\sim 20$ km ($\sim$ 1/2 ion-inertial length), were downstream $(\sim 1\, \mathrm{R_{E}})$ of the quasi-parallel bow shock. See Supplementary material for the location of MMS with respect to nominal magnetopause~\citep{Shue1998JGR} and bowshock~\citep{Farris1994JGR}. The magnetosheath interval has a flow speed of $238\, \mathrm{km\, s^{-1}}$, a density of 22 cm$^{-3}$,  and a proton beta 4.5. The average magnetic field is $B_0 \sim 18$ nT with fluctuations $\delta b \sim 14$ nT, so that $\delta b / B_0 \sim 0.8$. The interval displays standard features of well-developed turbulence, as previously studied in detail~\citep{ParasharPRL18}.}}
 
{We compare the MMS observations with the results from 
a 2.5-dimensional, fully kinetic, particle-in-cell (PIC) simulation~\citep{Yang2017PRE}. The simulation has $8192^2$ grid
points, with systems size $L = 102.4\,d_{\mathrm{i}}$, $\beta_{\mathrm{p}}=\beta_{\mathrm{e}} = 0.1$, $m_{\mathrm{p}} / m_{\mathrm{e}} = 25$, $\delta B / B_0 = 1/5$. We emphasize that 
no attempt is made to align the simulation parameters with those of the magnetosheath. In fact, one may note that the parameters like plasma beta, magnetic fluctuation amplitude are rather different from the particular interval analyzed here, and magnetosheath conditions~\citep{Huang2017ApJL} in general.}

In this paper, we are interested in the statistics of pressure strain interaction Pi-D $\equiv \pid$, which represents the incompressive channel of energy transfer into heat. The computation of $D_{ij}$ requires computation of velocity derivatives. 
The small separation in tetrahedron formation allows us to 
employ a straightforward
variation of the 
the curlometer technique~\cite{Dunlop1988ASR}, enabling
evaluation of the 
velocity strain tensor. Several previous studies have found that the curlometer technique is usually accurate for MMS data in the magnetosheath~\citep[e.g.,][]{Graham2016GRL, Gershman2018PoP, Stawarz2019ApJL}, although for some particular events, such as near large spatial gradients, the method may not be satisfactory~\citep{Akhavan-Tafti2018GRL}. For this particular interval, however, we find a reasonable agreement between the FPI and curlometer current (see Supplementary material). {The small elongation $(E \sim 0.3)$ and planarity $(P \sim 0.4)$ parameter values of the MMS tetrahedron configuration indicate adequate spatial coverage of the fluctuations~\citep{Robert1998ISSI}, so that one expects that the results are reliable.} The pressure tensor is averaged over the four MMS spacecraft. {The different temporal cadence of the MMS/FPI electron and ion measurements might, in principle, affect the comparison of the heating channels. However, we have found that the following results remain qualitatively unchanged, when performed with the electron data resampled to ion cadence (see Supplementary material).}

The proton and electron Pi-D, normalized by their rms fluctuations, are shown in Fig.~\ref{fig:timeseries}, along with normalized current density. {The intermittency of Pi-D is evident in the burstiness of these signals, with enhanced values concentrated in thin, sheet-like structures, occurring \textit{near} enhanced current density values.}

\begin{figure}
    \centering
    \includegraphics[width=0.7\columnwidth]{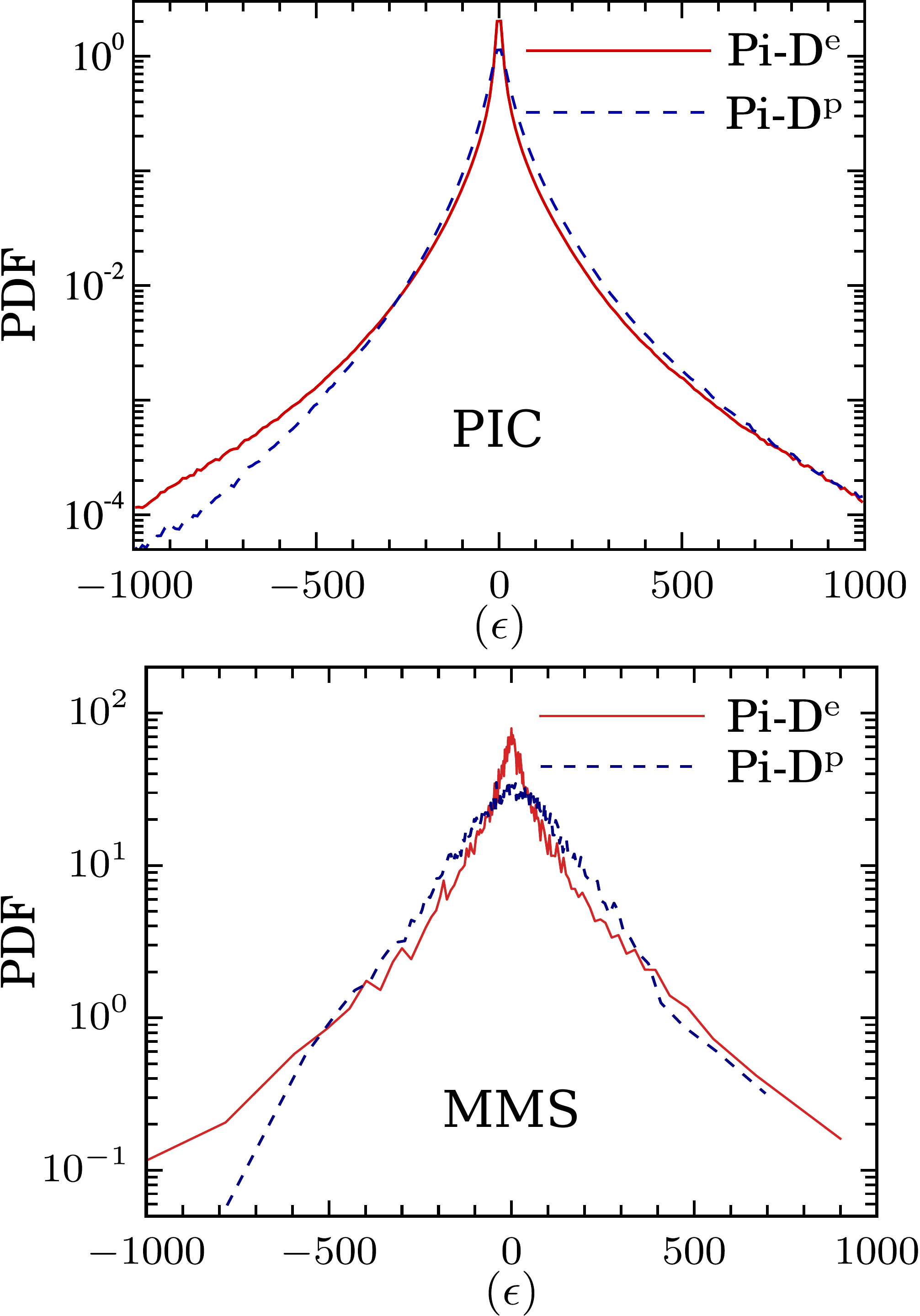}
    \caption{Probability distribution functions of Pi-D for protons (red solid line) as well as electrons (blue, dashed line) in (top) PIC simulations and (bottom) the magnetosheath from MMS data. The Pi-D values are normalized to the estimate of 
    large-scale decay rate $\epsilon$ (see text.)
    A tendency for protons to have slightly larger Pi-D can be seen. A slight preference for having higher positive tails is clear for both species.}
    \label{fig:pdfs}
\end{figure}

{We emphasize that Pi-D is a signed quantity in collisionless plasmas, as energy may be transferred into or out of the collective fluid motion. While pointwise these quantities are not sign-definite, the expectation is that when there is net dissipation and heating, the appropriate sign indicating net transfer into random motions will be favored. In contrast, in the case of viscous dissipation in collisional media, the Pi-D is positive definite by construction.} Nevertheless, the 
computed
mean value for Pi-D over the MMS interval is,
for protons, 
$ \langle -\Pi_{ij} D_{ij} \rangle = 4.8 \times 10^{-13}\, \mathrm{J\,m^{-3}\,s^{-1}}$, and, for electrons,
$4.5 \times 10^{-13}\,\mathrm{J\,m^{-3}\,s^{-1}}$.
This indicates 
a net transfer of energy from turbulence 
into random internal degrees of freedom during this interval. 
\begin{table}
	\caption{Turbulent heating measures from estimated evaluations at different scales.}
	\label{tab:pid}
	%\begin{center}
	\begin{tabularx}{\linewidth}{X X X X}
		\hline \hline
		$\epsilon_{\mathrm{von\,Karman}}$ &
		$\epsilon_{\mathrm{inertial}}$&
		$\langle$Pi-D$^{p}\rangle$ & $\langle$Pi-D$^{e}\rangle$ 
		\\
		$(\mathrm{J\,m^{-3}\,s^{-1}})$ & $(\mathrm{J\,m^{-3}\,s^{-1}})$ &
		$(\mathrm{J\,m^{-3}\,s^{-1}})$ &
		$(\mathrm{J\,m^{-3}\,s^{-1}})$ \\
		\hline
		 $(12.8 $ & $(9.4 $ & $ (5 $ & $ (4 $ \\
		 $ \pm 0.4) \times 10^{-14}$ & $ \pm 0.3) \times 10^{-14}$ & $ \pm 2) \times 10^{-13}$ & $ \pm 1) \times 10^{-13}$ \\
		\hline \hline
	\end{tabularx}
	%\end{center}
\end{table}

\begin{figure}
    \centering
    \includegraphics[width=\linewidth]{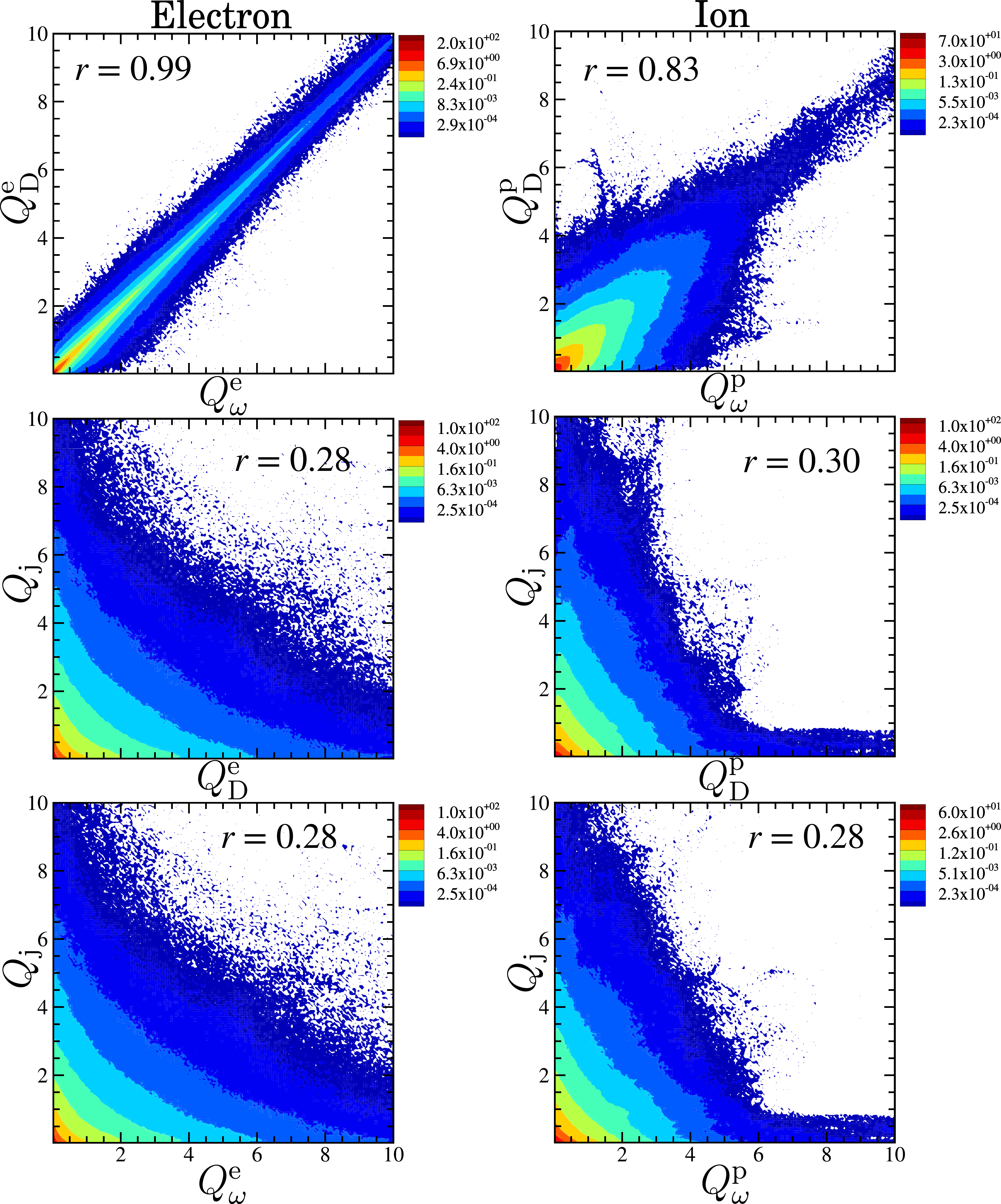}
    \caption{Joint probability distribution function of the normalized second invariants, $Q_{\omega} = (1/4){\boldsymbol{\omega}}^2/\langle{\boldsymbol{\omega}}^2\rangle$, $Q_{\mathrm{D}} = (1/4)\,D_{ij}  D_{ij}/\langle D_{ij} D_{ij} \rangle$, and $Q_{\mathrm{j}} = (1/4){\mathbf{j}}^2/\langle {\mathbf{j}}^2 \rangle$ for electrons (left column) and protons (right column) from PIC data~\cite{Yang2017PoP}. {Pearson correlation coefficient $(r)$ is shown for each panel.}}
    \label{fig:jpdf_pic}
\end{figure}

\begin{figure}
    \centering
    \includegraphics[width=\linewidth]{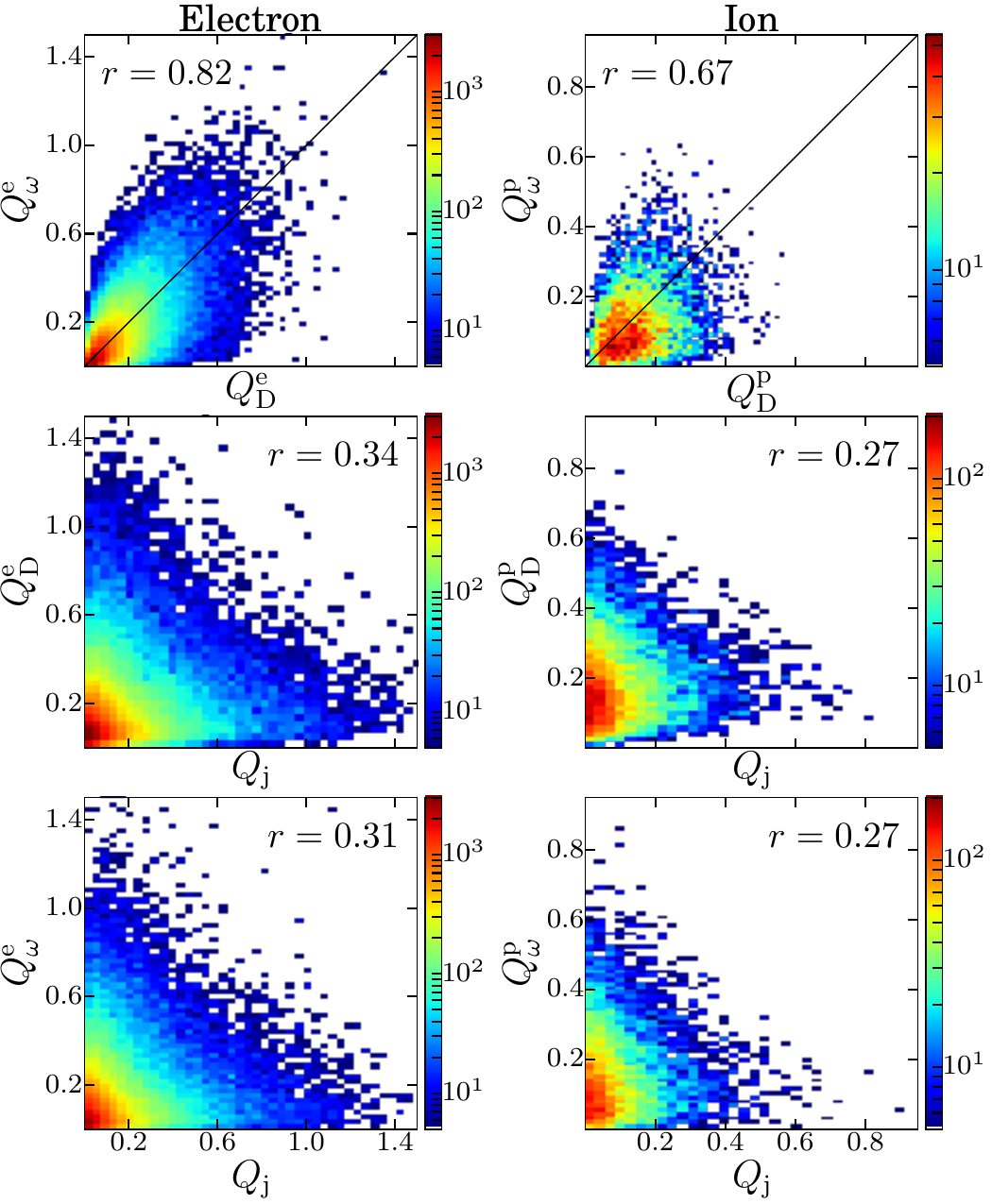}
    \caption{Same as Fig.~\ref{fig:jpdf_pic}, but obtained from MMS observations.}
    \label{fig:jpdf_mms}
\end{figure}

To establish a clear connection of the collisionless dissipation measure, Pi-D, with the fluid-scale energy transfer rates, we compare the net (averaged) Pi-D with the MHD measures of decay rate. We evaluate the von K\'arm\'an law and third-order law, in a manner similar to that performed in \citep{Bandyopadhyay2018bApJ}. Table~\ref{tab:pid} reports the approximate values of energy-transfer rate, obtained from the three constructs, at  different ranges of scale, and the proton and electron Pi-D averages.

There is a reasonable level of agreement among the three measures, indicating an approximate validity of the general scheme of fluid-scale energy cascade, eventually heating the protons and electrons. Variability is likely due to poor statistics, anisotropy of the turbulence, and the possibility of coupling with the compressive channel of energy conversion. {A detailed statistical survey with many MMS intervals would help to clarify some of these issues.}

The average rate of incompressive heating as well as 
associated fluctuations 
may also be seen by examining
the probability distribution functions (PDFs) of Pi-D for both species, illustrated in Fig.~\ref{fig:pdfs}. {To make a more direct comparison of the simulation and observation, we normalize the Pi-D values to the global decay rates, $\epsilon$. In simulations, this is evaluated simply by computing the rate of change of total (magnetic+flow) energy, and for MMS data the von K\'arm\'an estimate (Table~\ref{tab:pid}) is used.} The curves are highly non-Gaussian, providing an additional indication of the intermittent distribution of Pi-D. {The total kurtosis, defined for variable $x$ as $\kappa = \langle (x-\langle x \rangle)^{4} \rangle / {\langle (x-\langle x \rangle)^{2} \rangle}^{2}$, is $24.6$ for the ion Pi-D and $41.6$ for the electron Pi-D.} The high values of kurtosis reflect the strong intermittency in these variables.

{The burstiness of Pi-D, as seen in Fig.~\ref{fig:timeseries}, 
suggests correlations with current density, as well as 
other physical quantities, such as vorticity $(\boldsymbol{\omega} = \boldsymbol{\nabla} \times \m{u})$ and symmetric velocity strain $(D_{ij})$, which often exhibit similar non-uniform distribution in plasmas. We can examine such possibilities by 
studying the spatial concentration of Pi-D in comparison with $D_{ij}$, $\boldsymbol{\omega}$ , and $\m{j}$}. We normalize the three second-order invariants as
$Q_{\omega} = (1/4){\boldsymbol{\omega}}^2/\langle{\boldsymbol{\omega}}^2\rangle$, $Q_{\mathrm{D}} = (1/4)\,D_{ij}\, D_{ij}/\langle D_{ij}\,D_{ij} \rangle$, and $Q_{\mathrm{j}} = (1/4){\mathbf{j}}^2/\langle {\mathbf{j}}^2 \rangle$. The invariant $Q_{\omega}$ represents rotation, $Q_{\mathrm{D}}$ corresponds to straining motions, and $Q_{\mathrm{j}}$ is related to magnetic gradients. All of them can interact with one another. {To explore the spatial correlation of these processes, we show the joint PDF  of the normalized second invariants for each species in Fig.~\ref{fig:jpdf_pic} and Fig.~\ref{fig:jpdf_mms}. To obtain a quantitative assessment, we report the Pearson linear correlation coefficient for each pair of invariants.} 

{From the top two panels in Fig.~\ref{fig:jpdf_pic} and Fig.~\ref{fig:jpdf_mms}, the joint PDFs of $Q_{\omega}$ and $Q_{\mathrm{D}}$ are dominated by a population near the $Q_{\omega} = Q_{\mathrm{D}}$ line, demonstrating a strong spatial correlation between the two quantities.
This strong correlation, found here in plasma turbulence, resembles similar results in 
hydrodynamic turbulence 
in regimes in which vorticity is sheet-like rather than tube-like~\citep{jimenez1993JFM, Blackburn1996JFM}. Further, similar to what is observed from the plasma simulations (Fig.~\ref{fig:jpdf_pic}), the positive correlation in MMS observation is very prominent for electrons, but somewhat weaker in the case of protons (Fig.~\ref{fig:jpdf_mms}). Although, the better correlation for the case of the electrons in MMS data may be a result of larger statistical sample and better accuracy due to a higher temporal resolution. 
We note that, if this result were to be established as accurate, it would imply that electron vorticity has a very strong tendency to appear in 
sheets. The joint PDFs of $Q_{\mathrm{j}}$ versus $Q_{\mathrm{D}}$, in contrast, are spread broadly, with low correlation coefficients, indicating weak pointwise correlation between these quantities. Similarly, the joint PDF of $Q_{\mathrm{j}}$ versus $Q_{\omega}$ exhibit weak correlation with small correlation coefficient both for PIC and MMS case. Therefore, the vorticity and traceless strain-rate tensor do not correlate pointwise with current density, but are slightly offset in space.}

\begin{figure}
    \centering
    \includegraphics[width=\linewidth]{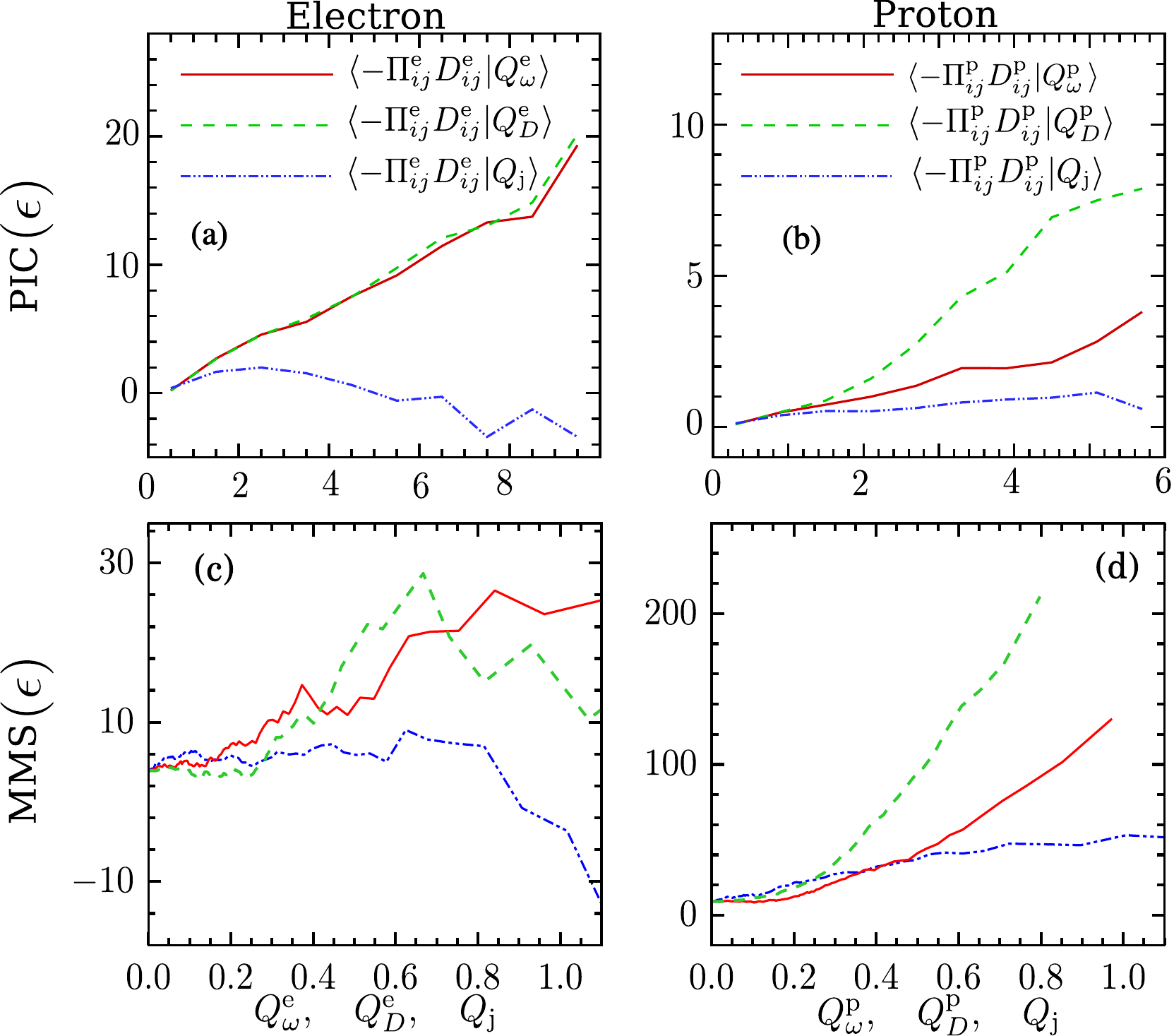}
    \caption{Conditional averages of the electron and proton Pi-D term from PIC simulation (top); and the same from MMS data (bottom). The Pi-D values are normalized to large-scale decay rates $\epsilon$ (see text).}
    \label{fig:condav}
\end{figure}

To quantify the spatial correlation between Pi-D and symmetric velocity stress, vorticity, and current density, we compute the  conditional averages of Pi-D with these quantities. Figure~\ref{fig:condav}
plots the conditional averages of $-\Pi_{ij} D_{ ij}$ , separately for protons
and electrons. The conditions are based on values of the second tensor 
invariants $Q_\mathrm{D}$, $Q_{\omega}$, and $Q_\mathrm{j}$. For example, to compute 
$\langle - \Pi_{ij}^{e} D_{ij}^{e} |Q_\mathrm{j} \rangle $, one calculates the 
average of the electron Pi-D including only the values occurring at
times when the mean-square total electric current
density ($Q_\mathrm{j}$) exceeds a selected threshold. The figure indicates
that, for both electrons and protons, elevated levels of $− \Pi_{ij} D_{ij}$
are found in regions with enhanced vorticity and in regions of enhanced
symmetric stress, consistent
with earlier reports~\cite{Yang2017PRE, Yang2017PoP}. In contrast, the averages of Pi-D conditioned
on total current density remain fairly constant for protons, and
slightly decrease for electrons. The values of Pi-D for
protons are even more elevated in regions of large symmetric
stress than in regions of large (mean-square) vorticity. The similarity to 
the analogous
results obtained from kinetic simulations in \cite{Yang2017PoF} 
is once again striking, 
suggesting that the properties reported here 
are fundamental to weakly-collisional plasmas,
and not particular to a specific set of parameters.

{Although the results are in qualitative agreement, the range of values of some variables 
are sometimes quite 
different in the two systems, especially for the protons in Fig.~\ref{fig:condav}. 
For example, the range of normalized  proton Pi-D values in Fig.~\ref{fig:condav}, are different. 
Such 
disparity in the two systems 
is likely attributable 
to the artificial simulation 
mass ratio, 
different scale separations and system sizes, and differences in 
large-scale driving mechanisms.
}

%\section{Discussion}
In this paper, we have presented a statistical characterization,
of the direct pathways to production of internal energy 
in collisioness plasma turbulence. 
In particular we employ
MMS observations in the terrestrial magnetosheath
to quantify production of internal energy
through the pressure-strain interaction,
namely the $-\Pi_{ij} D_{ij}$ term.
Previous studies have computed
Pi-D in individual events, such as current sheets~\citep{Chasapis2018ApJL}.
The present study is the first one we are aware of that 
has derived statistical distributions
of pressure strain from a large continuous dataset. 
It is important to recall  that the statistics of pressure-strain provide a direct quantitative measure of internal energy production without the usual restrictions inherent in selection in advance of a particular wave-mode or mechanism. 
In this way  the present results provide insights into dissipation that are potentially more general than those based on specific mechanisms. Direct comparison between statistics obtained from simulation and from MMS observations show a remarkable qualitative level of agreement. {Note that additional supporting  analysis, including 
an additional MMS interval, is provided as 
Supplementary material, with conclusion consistent with those shown here.}

The scale-to-scale energy transfer process has been well-studied
in the energy-containing and inertial range \cite{Verma04,Coburn2014PRS, Bandyopadhyay2018bApJ}, but the energy conversion processes in the kinetic ranges are not understood. 
The results presented in this paper provide a step towards that direction,
suggesting correlations and channels of energy conversion that with further study may provide broader insights into these essential plasma physics processes.

\begin{acknowledgments}
This research was supported in part by the MMS Theory and Modeling team grant 
under NASA grant NNX14AC39 and by NASA
Heliophysics SRT grant NNX17AB79G.
We are
grateful to the MMS instrument teams for cooperation
and collaboration in preparing the data. The data
used in this analysis are Level 2 FIELDS and FPI data
products, in cooperation with the instrument teams
and in accordance their guidelines. All MMS data are
available at \url{https://lasp.colorado.edu/mms/sdc/}. The Wind data, shifted to
Earth’s bow-shock nose, can be found at \url{https://omniweb.gsfc.nasa.gov/}. The authors thank the Wind team for the magnetic field and proton moment dataset.
\end{acknowledgments}

\section{supplemental material}
In this supplemental material, we present additional materials which support the content of the main letter.

\subsection{MMS Location}
\begin{figure}[ht!]
	\centering
	\includegraphics[width=0.76\linewidth]{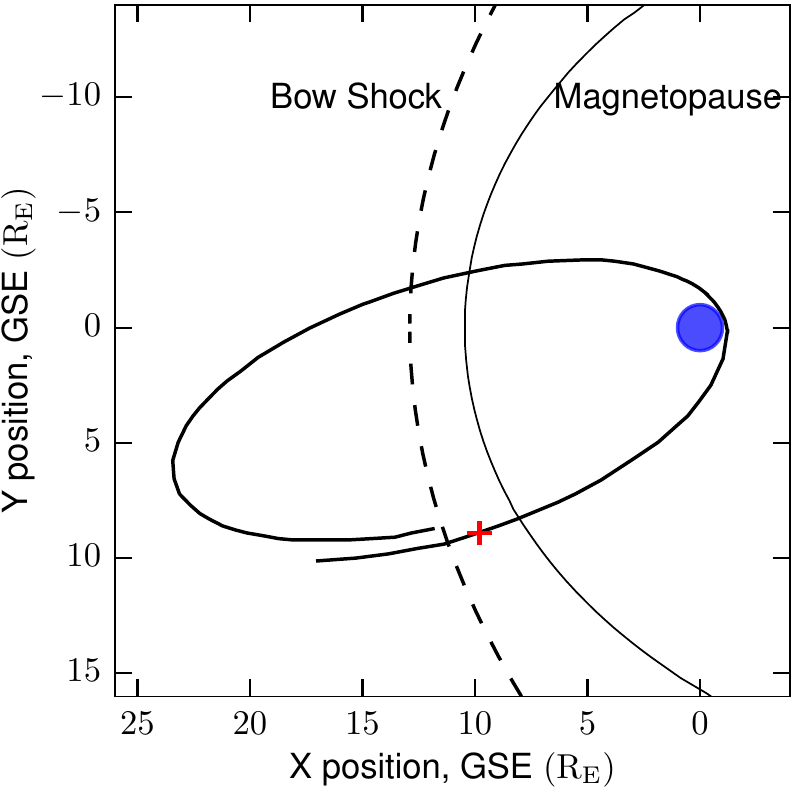}
	\caption{MMS orbit in GSE coordinates during the magnetosheath interval analyzed in the letter. The lengths are in units of Earth Radii. Red “+” symbol
		marks location of MMS during the intervals examined here (26 December 2017, 0:36:14). Nominal locations of the bow shock and the
		magnetopause are also shown.}
	\label{fig:orb}
\end{figure}
The nature of turbulent fluctuations may be different depending on the location of the spacecraft in the magnetosheath~\citep{Huang2017ApJL}. To provide context, we depict the location of the spacecraft during the analyzed interval in Figure 2, along with nominal locations of the magnetopause and the bow shock. The specific MMS orbit is also shown for relevance.  The
geocentric solar ecliptic (GSE) coordinate system is used, in which the XY-plane is defined by the Earth mean
ecliptic of date and the +X-axis is defined by the Earth-Sun vector. The location of the bow shock was estimated using the model of \citep{Farris1994JGR}, and the location of the
magnetopause was obtained from the MMS SDC, which employs the model of \citep{Shue1998JGR}.

%\clearpage

\subsection{The Curlometer Technique}
\begin{figure}
	\centering
	\includegraphics[width=0.6 \linewidth]{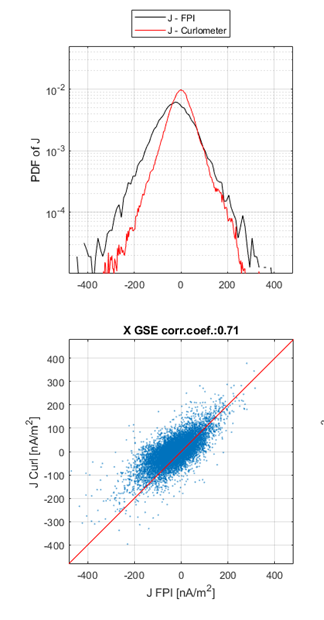}
	\caption{Probability Distribution Function (PDF) and scatter plot of current density (X component) calculated from the curlometer technique and FPI. }
	\label{fig:curlo}
\end{figure}

To check the accuracy of the curlometer method, Fig.~\ref{fig:curlo} shows the comparison of X component of current density measured using the curlometer technique~\citep{Dunlop1988ASR}, $\mathbf{j} = (1/\mu_0)\mathbf{\nabla}\times\mathbf{b}$, and that measured by computing the difference of ion and electron velocities, $\mathbf{j} = n_{\mathrm{e}}(\mathbf{u}_{\mathrm{p}} - \mathbf{u}_{\mathrm{e}})$, for the magnetosheath interval (26 December 2017, 06:12:43 - 06:52:23 UTC) analyzed in the letter. Similar agreements are found for the Y and Z component (not shown), with the Pearson linear correlation coefficient value of 0.81 and 0.72, respectively. The results are not perfect but they are satisfactory and almost certainly the state of the art in current measurements in space data sets.

%\clearpage

\subsection{Additional Supporting Analysis}
Here, we reevaluate some of the results presented in the main text for another \textit{long} $(\approx 20 \,\mathrm{min})$ magneotsheath interval sampled by MMS. %The location of the MMS spacecraft, along with the nominal position of the bow shock and the magnetopause for this interval are shown in Fig.~\ref{fig:orb2}. 
The period lasts from 07:21:54 to 07:48:01 UTC, on 21 December 2017. At this time, the interplanetary solar wind had an average magnetic field of $ 4.7\,\mathrm{nT}$, flow speed 400 $\mathrm{km s^{-1}}$, density  4.4 cm$^{-3}$, and proton temperature 19200 K. The MMS spacecraft, separated by $\sim 20$ km ($\sim$ 1/2 ion-inertial length), were downstream of the quasi-parallel bow shock. The magnetosheath interval has a mean flow speed of 100 $\mathrm{km s^{-1}}$, a density of 21 cm$^{-3}$,  and a proton beta 4.7. The average magnetic field is $B_0 \sim 10$ nT, and the level of fluctuations $\delta b \sim 20$ nT, so that $\delta b / B_0 \sim 2$.

\begin{figure}
	\centering
	\includegraphics[width=\columnwidth]{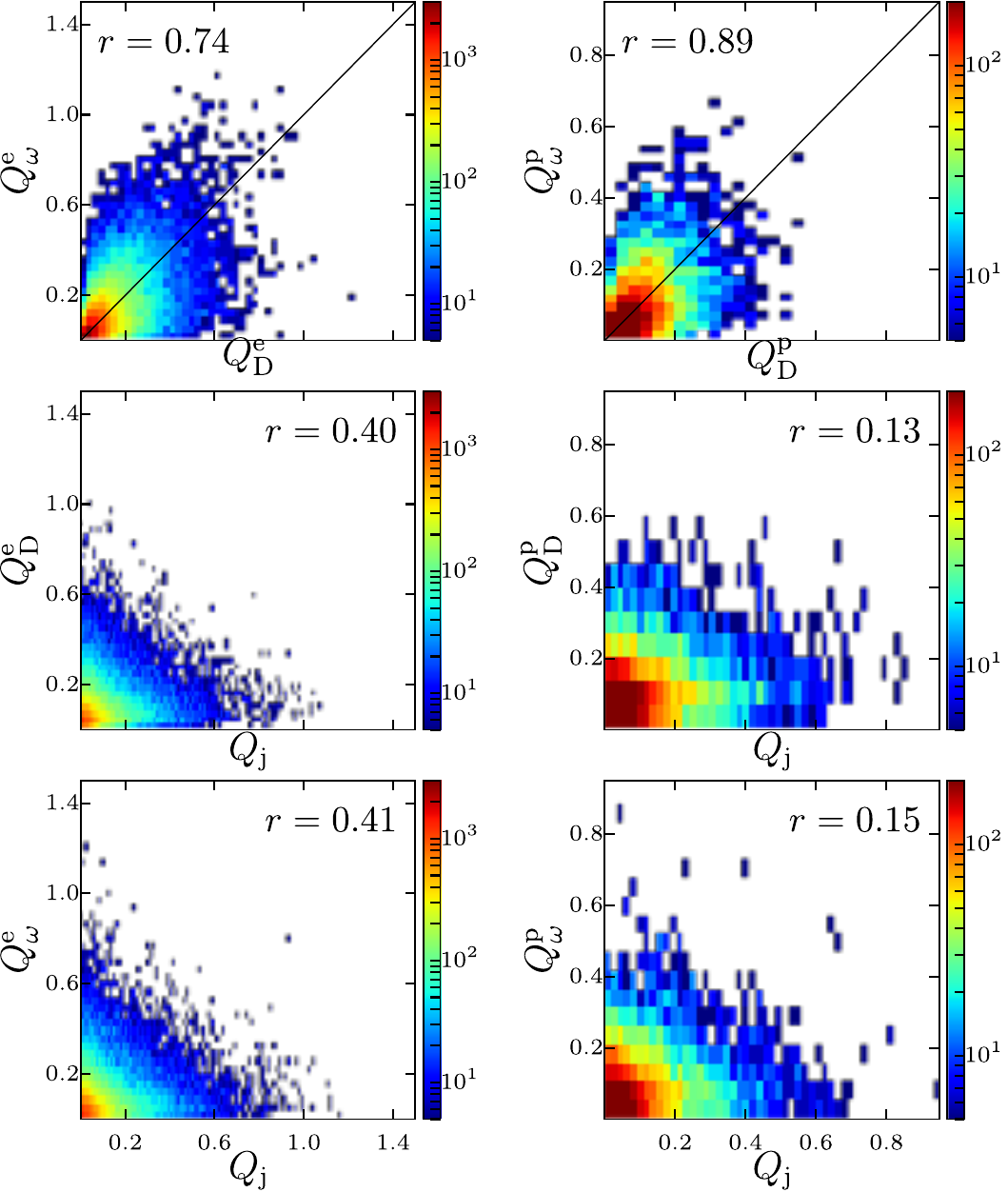}
	\caption{Joint probability distribution function of the normalized second invariants of rotation-rate, traceless strain-rate tensors, and current density, i.e.,$Q_{\omega} = (1/4){\boldsymbol{\omega}}^2/\langle{\boldsymbol{\omega}}^2\rangle$, $Q_{\mathrm{D}} = (1/4)\,D_{ij}  D_{ij}/\langle D_{ij} D_{ij} \rangle$, and $Q_{\mathrm{j}} = (1/4){\mathbf{j}}^2/\langle {\mathbf{j}}^2 \rangle$ for electrons (left column) and protons (right column) for another MMS interval. Pearson's correlation coefficient are shown for each panel.}
	\label{fig:jpdfs}
\end{figure}

Figure~\ref{fig:jpdfs} shows the joint probability distributions for the
second-rank invariants of the
vorticity, current density and symmetric strain tensors for the electrons and 
for the protons. These are 
completely analogous to Fig. 4 in the main letter, except this figure is produced using the distinct interval mentioned above. 

\begin{figure}
	\centering
	\includegraphics[width=\linewidth]{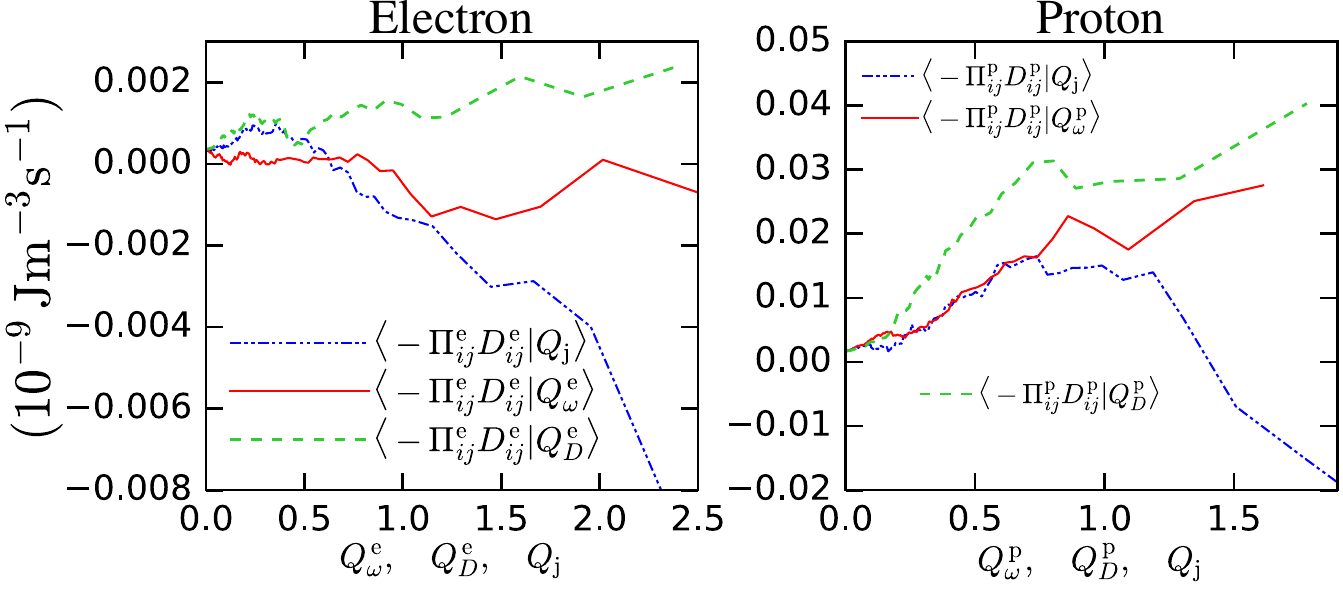}
	\caption{Conditional averages of the (left) electron Pi-D term and (right) proton Pi-D term from another MMS interval in the magnetosheath.}
	\label{fig:conds}
\end{figure}

Similarly, Fig.~\ref{fig:conds}  
shows the averages of the Pi-D terms 
conditioned on the same 
second order invariants; it is 
equivalent to Fig. 5 in the main letter, except now the figure is produced using the second interval.

We note that the results shown in 
Fig. \ref{fig:jpdfs} and Fig.~\ref{fig:conds}
are quite similar to the corresponding figures in the main text. Although this is in no way a fully exhaustive sampling, this similarity suggests that the behavior seen may be typical, at least for this parameter range of plasma 
turbulence. 

We are currently engaged in extending this study to include more available MMS intervals of this type. One main issue is  
that most of the available
intervals are much shorter than the 
present interval. A proper analysis of such a collection will certainly involve 
new issues, such as normalizing different quantities 
in the intervals so that the data can be compared and combined. It will also be interesting to extend these analyses to a broader class of samples 
from different regions in the near-Earth plasma environment, wherever MMS capabilities may allow, in order 
to further study variations
with plasma parameters and conditions. 
We defer such 
investigations to subsequent studies.

\subsection{Electron Cadence Data}
\begin{figure}[hb!]
	\centering
	\includegraphics[width=\linewidth]{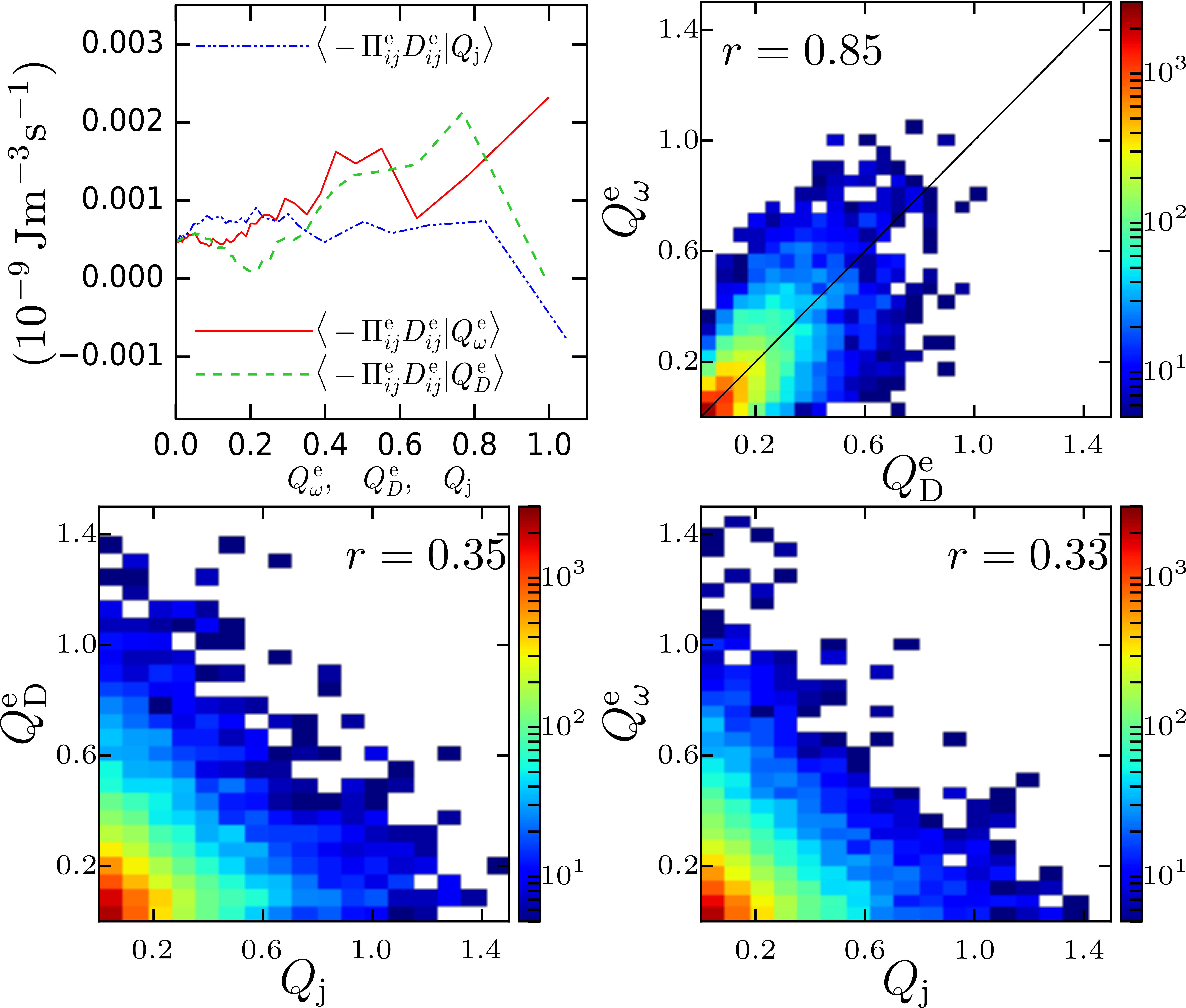}
	\caption{Some electron results reproduced with ion cadence.}
	\label{fig:e}
\end{figure}
The lower time resolution of ion data may affect the accuracy of the calculations. To get a rough idea, we resampled the electron data to the ion cadence and redid all the calculations. We find that this does not change the results qualitatively and the  conclusions remain the same. In Fig.~\ref{fig:e} we have included the equivalent of panel c) of Fig. 5 from the main paper, evaluated using electron data resampled to the ion cadence. Comparison with panel c) of Fig 5 in the paper confirms that there is little difference and the main result is qualitatively unchanged. We also show new panels for the electron correlations at the resampled cadences, in analogy to the left column of Fig. 4 in the main paper. Note that the correlation coefficients are almost the same. We can also verify that the qualitative shape of the correlations is unchanged.

%\bibliography{pid_stats,refs_riddhi,refs_WHM,AG,HL,MP,QZ}

\begin{thebibliography}{40}%
	\makeatletter
	\providecommand \@ifxundefined [1]{%
		\@ifx{#1\undefined}
	}%
	\providecommand \@ifnum [1]{%
		\ifnum #1\expandafter \@firstoftwo
		\else \expandafter \@secondoftwo
		\fi
	}%
	\providecommand \@ifx [1]{%
		\ifx #1\expandafter \@firstoftwo
		\else \expandafter \@secondoftwo
		\fi
	}%
	\providecommand \natexlab [1]{#1}%
	\providecommand \enquote  [1]{``#1''}%
	\providecommand \bibnamefont  [1]{#1}%
	\providecommand \bibfnamefont [1]{#1}%
	\providecommand \citenamefont [1]{#1}%
	\providecommand \href@noop [0]{\@secondoftwo}%
	\providecommand \href [0]{\begingroup \@sanitize@url \@href}%
	\providecommand \@href[1]{\@@startlink{#1}\@@href}%
	\providecommand \@@href[1]{\endgroup#1\@@endlink}%
	\providecommand \@sanitize@url [0]{\catcode `\\12\catcode `\$12\catcode
		`\&12\catcode `\#12\catcode `\^12\catcode `\_12\catcode `\%12\relax}%
	\providecommand \@@startlink[1]{}%
	\providecommand \@@endlink[0]{}%
	\providecommand \url  [0]{\begingroup\@sanitize@url \@url }%
	\providecommand \@url [1]{\endgroup\@href {#1}{\urlprefix }}%
	\providecommand \urlprefix  [0]{URL }%
	\providecommand \Eprint [0]{\href }%
	\providecommand \doibase [0]{http://dx.doi.org/}%
	\providecommand \selectlanguage [0]{\@gobble}%
	\providecommand \bibinfo  [0]{\@secondoftwo}%
	\providecommand \bibfield  [0]{\@secondoftwo}%
	\providecommand \translation [1]{[#1]}%
	\providecommand \BibitemOpen [0]{}%
	\providecommand \bibitemStop [0]{}%
	\providecommand \bibitemNoStop [0]{.\EOS\space}%
	\providecommand \EOS [0]{\spacefactor3000\relax}%
	\providecommand \BibitemShut  [1]{\csname bibitem#1\endcsname}%
	\let\auto@bib@innerbib\@empty
	%</preamble>
	\bibitem [{\citenamefont {Quataert}(2003)}]{QuataertAN03}%
	\BibitemOpen
	\bibfield  {author} {\bibinfo {author} {\bibfnamefont {E.}~\bibnamefont
			{Quataert}},\ }\href@noop {} {\bibfield  {journal} {\bibinfo  {journal}
			{Astronomische Nachrichten}\ }\textbf {\bibinfo {volume} {324}},\ \bibinfo
		{pages} {435} (\bibinfo {year} {2003})}\BibitemShut {NoStop}%
	\bibitem [{\citenamefont {Marsch}(2006)}]{Marsch06}%
	\BibitemOpen
	\bibfield  {author} {\bibinfo {author} {\bibfnamefont {E.}~\bibnamefont
			{Marsch}},\ }\href {http://www.livingreviews.org/lrsp-2006-1} {\bibfield
		{journal} {\bibinfo  {journal} {Living Reviews of Solar Physics}\ }\textbf
		{\bibinfo {volume} {3}} (\bibinfo {year} {2006})}\BibitemShut {NoStop}%
	\bibitem [{\citenamefont {Wang}\ and\ \citenamefont
		{Richardson}(2001)}]{WangJGR01}%
	\BibitemOpen
	\bibfield  {author} {\bibinfo {author} {\bibfnamefont {C.}~\bibnamefont
			{Wang}}\ and\ \bibinfo {author} {\bibfnamefont {J.}~\bibnamefont
			{Richardson}},\ }\href@noop {} {\bibfield  {journal} {\bibinfo  {journal}
			{Journal of Geophysical Research}\ }\textbf {\bibinfo {volume} {106}},\
		\bibinfo {pages} {29401} (\bibinfo {year} {2001})}\BibitemShut {NoStop}%
	\bibitem [{\citenamefont {Burch}\ \emph {et~al.}(2016)\citenamefont {Burch},
		\citenamefont {Moore}, \citenamefont {Torbert},\ and\ \citenamefont
		{Giles}}]{Burch2016SSR}%
	\BibitemOpen
	\bibfield  {author} {\bibinfo {author} {\bibfnamefont {J.~L.}\ \bibnamefont
			{Burch}}, \bibinfo {author} {\bibfnamefont {T.~E.}\ \bibnamefont {Moore}},
		\bibinfo {author} {\bibfnamefont {R.~B.}\ \bibnamefont {Torbert}}, \ and\
		\bibinfo {author} {\bibfnamefont {B.~L.}\ \bibnamefont {Giles}},\ }\href
	{\doibase 10.1007/s11214-015-0164-9} {\bibfield  {journal} {\bibinfo
			{journal} {Space Science Reviews}\ }\textbf {\bibinfo {volume} {199}},\
		\bibinfo {pages} {5} (\bibinfo {year} {2016})}\BibitemShut {NoStop}%
	\bibitem [{\citenamefont {Pollock}\ \emph {et~al.}(2016)\citenamefont
		{Pollock}, \citenamefont {Moore}, \citenamefont {Jacques}, \citenamefont
		{Burch}, \citenamefont {Gliese}, \citenamefont {Saito}, \citenamefont
		{Omoto}, \citenamefont {Avanov}, \citenamefont {Barrie}, \citenamefont
		{Coffey}, \citenamefont {Dorelli}, \citenamefont {Gershman}, \citenamefont
		{Giles}, \citenamefont {Rosnack}, \citenamefont {Salo}, \citenamefont
		{Yokota}, \citenamefont {Adrian}, \citenamefont {Aoustin}, \citenamefont
		{Auletti}, \citenamefont {Aung}, \citenamefont {Bigio}, \citenamefont {Cao},
		\citenamefont {Chandler}, \citenamefont {Chornay}, \citenamefont {Christian},
		\citenamefont {Clark}, \citenamefont {Collinson}, \citenamefont {Corris},
		\citenamefont {De Los Santos}, \citenamefont {Devlin}, \citenamefont
		{Diaz}, \citenamefont {Dickerson}, \citenamefont {Dickson}, \citenamefont
		{Diekmann}, \citenamefont {Diggs}, \citenamefont {Duncan}, \citenamefont
		{Figueroa-Vinas}, \citenamefont {Firman}, \citenamefont {Freeman},
		\citenamefont {Galassi}, \citenamefont {Garcia}, \citenamefont {Goodhart},
		\citenamefont {Guererro}, \citenamefont {Hageman}, \citenamefont {Hanley},
		\citenamefont {Hemminger}, \citenamefont {Holland}, \citenamefont {Hutchins},
		\citenamefont {James}, \citenamefont {Jones}, \citenamefont {Kreisler},
		\citenamefont {Kujawski}, \citenamefont {Lavu}, \citenamefont {Lobell},
		\citenamefont {LeCompte}, \citenamefont {Lukemire}, \citenamefont
		{MacDonald}, \citenamefont {Mariano}, \citenamefont {Mukai}, \citenamefont
		{Narayanan}, \citenamefont {Nguyan}, \citenamefont {Onizuka}, \citenamefont
		{Paterson}, \citenamefont {Persyn}, \citenamefont {Piepgrass}, \citenamefont
		{Cheney}, \citenamefont {Rager}, \citenamefont {Raghuram}, \citenamefont
		{Ramil}, \citenamefont {Reichenthal}, \citenamefont {Rodriguez},
		\citenamefont {Rouzaud}, \citenamefont {Rucker}, \citenamefont {Saito},
		\citenamefont {Samara}, \citenamefont {Sauvaud}, \citenamefont {Schuster},
		\citenamefont {Shappirio}, \citenamefont {Shelton}, \citenamefont {Sher},
		\citenamefont {Smith}, \citenamefont {Smith}, \citenamefont {Smith},
		\citenamefont {Steinfeld}, \citenamefont {Szymkiewicz}, \citenamefont
		{Tanimoto}, \citenamefont {Taylor}, \citenamefont {Tucker}, \citenamefont
		{Tull}, \citenamefont {Uhl}, \citenamefont {Vloet}, \citenamefont {Walpole},
		\citenamefont {Weidner}, \citenamefont {White}, \citenamefont {Winkert},
		\citenamefont {Yeh},\ and\ \citenamefont {Zeuch}}]{Pollock2016SSR}%
	\BibitemOpen
	\bibfield  {author} {\bibinfo {author} {\bibfnamefont {C.}~\bibnamefont
			{Pollock}}, \bibinfo {author} {\bibfnamefont {T.}~\bibnamefont {Moore}},
		\bibinfo {author} {\bibfnamefont {A.}~\bibnamefont {Jacques}}, \bibinfo
		{author} {\bibfnamefont {J.}~\bibnamefont {Burch}}, \bibinfo {author}
		{\bibfnamefont {U.}~\bibnamefont {Gliese}}, \bibinfo {author} {\bibfnamefont
			{Y.}~\bibnamefont {Saito}}, \bibinfo {author} {\bibfnamefont
			{T.}~\bibnamefont {Omoto}}, \bibinfo {author} {\bibfnamefont
			{L.}~\bibnamefont {Avanov}}, \bibinfo {author} {\bibfnamefont
			{A.}~\bibnamefont {Barrie}}, \bibinfo {author} {\bibfnamefont
			{V.}~\bibnamefont {Coffey}}, \bibinfo {author} {\bibfnamefont
			{J.}~\bibnamefont {Dorelli}}, \bibinfo {author} {\bibfnamefont
			{D.}~\bibnamefont {Gershman}}, \bibinfo {author} {\bibfnamefont
			{B.}~\bibnamefont {Giles}}, \bibinfo {author} {\bibfnamefont
			{T.}~\bibnamefont {Rosnack}}, \bibinfo {author} {\bibfnamefont
			{C.}~\bibnamefont {Salo}}, \bibinfo {author} {\bibfnamefont {S.}~\bibnamefont
			{Yokota}}, \bibinfo {author} {\bibfnamefont {M.}~\bibnamefont {Adrian}},
		\bibinfo {author} {\bibfnamefont {C.}~\bibnamefont {Aoustin}}, \bibinfo
		{author} {\bibfnamefont {C.}~\bibnamefont {Auletti}}, \bibinfo {author}
		{\bibfnamefont {S.}~\bibnamefont {Aung}}, \bibinfo {author} {\bibfnamefont
			{V.}~\bibnamefont {Bigio}}, \bibinfo {author} {\bibfnamefont
			{N.}~\bibnamefont {Cao}}, \bibinfo {author} {\bibfnamefont {M.}~\bibnamefont
			{Chandler}}, \bibinfo {author} {\bibfnamefont {D.}~\bibnamefont {Chornay}},
		\bibinfo {author} {\bibfnamefont {K.}~\bibnamefont {Christian}}, \bibinfo
		{author} {\bibfnamefont {G.}~\bibnamefont {Clark}}, \bibinfo {author}
		{\bibfnamefont {G.}~\bibnamefont {Collinson}}, \bibinfo {author}
		{\bibfnamefont {T.}~\bibnamefont {Corris}}, \bibinfo {author} {\bibfnamefont
			{A.}~\bibnamefont {De Los Santos}}, \bibinfo {author} {\bibfnamefont
			{R.}~\bibnamefont {Devlin}}, \bibinfo {author} {\bibfnamefont
			{T.}~\bibnamefont {Diaz}}, \bibinfo {author} {\bibfnamefont {T.}~\bibnamefont
			{Dickerson}}, \bibinfo {author} {\bibfnamefont {C.}~\bibnamefont {Dickson}},
		\bibinfo {author} {\bibfnamefont {A.}~\bibnamefont {Diekmann}}, \bibinfo
		{author} {\bibfnamefont {F.}~\bibnamefont {Diggs}}, \bibinfo {author}
		{\bibfnamefont {C.}~\bibnamefont {Duncan}}, \bibinfo {author} {\bibfnamefont
			{A.}~\bibnamefont {Figueroa-Vinas}}, \bibinfo {author} {\bibfnamefont
			{C.}~\bibnamefont {Firman}}, \bibinfo {author} {\bibfnamefont
			{M.}~\bibnamefont {Freeman}}, \bibinfo {author} {\bibfnamefont
			{N.}~\bibnamefont {Galassi}}, \bibinfo {author} {\bibfnamefont
			{K.}~\bibnamefont {Garcia}}, \bibinfo {author} {\bibfnamefont
			{G.}~\bibnamefont {Goodhart}}, \bibinfo {author} {\bibfnamefont
			{D.}~\bibnamefont {Guererro}}, \bibinfo {author} {\bibfnamefont
			{J.}~\bibnamefont {Hageman}}, \bibinfo {author} {\bibfnamefont
			{J.}~\bibnamefont {Hanley}}, \bibinfo {author} {\bibfnamefont
			{E.}~\bibnamefont {Hemminger}}, \bibinfo {author} {\bibfnamefont
			{M.}~\bibnamefont {Holland}}, \bibinfo {author} {\bibfnamefont
			{M.}~\bibnamefont {Hutchins}}, \bibinfo {author} {\bibfnamefont
			{T.}~\bibnamefont {James}}, \bibinfo {author} {\bibfnamefont
			{W.}~\bibnamefont {Jones}}, \bibinfo {author} {\bibfnamefont
			{S.}~\bibnamefont {Kreisler}}, \bibinfo {author} {\bibfnamefont
			{J.}~\bibnamefont {Kujawski}}, \bibinfo {author} {\bibfnamefont
			{V.}~\bibnamefont {Lavu}}, \bibinfo {author} {\bibfnamefont {J.}~\bibnamefont
			{Lobell}}, \bibinfo {author} {\bibfnamefont {E.}~\bibnamefont {LeCompte}},
		\bibinfo {author} {\bibfnamefont {A.}~\bibnamefont {Lukemire}}, \bibinfo
		{author} {\bibfnamefont {E.}~\bibnamefont {MacDonald}}, \bibinfo {author}
		{\bibfnamefont {A.}~\bibnamefont {Mariano}}, \bibinfo {author} {\bibfnamefont
			{T.}~\bibnamefont {Mukai}}, \bibinfo {author} {\bibfnamefont
			{K.}~\bibnamefont {Narayanan}}, \bibinfo {author} {\bibfnamefont
			{Q.}~\bibnamefont {Nguyan}}, \bibinfo {author} {\bibfnamefont
			{M.}~\bibnamefont {Onizuka}}, \bibinfo {author} {\bibfnamefont
			{W.}~\bibnamefont {Paterson}}, \bibinfo {author} {\bibfnamefont
			{S.}~\bibnamefont {Persyn}}, \bibinfo {author} {\bibfnamefont
			{B.}~\bibnamefont {Piepgrass}}, \bibinfo {author} {\bibfnamefont
			{F.}~\bibnamefont {Cheney}}, \bibinfo {author} {\bibfnamefont
			{A.}~\bibnamefont {Rager}}, \bibinfo {author} {\bibfnamefont
			{T.}~\bibnamefont {Raghuram}}, \bibinfo {author} {\bibfnamefont
			{A.}~\bibnamefont {Ramil}}, \bibinfo {author} {\bibfnamefont
			{L.}~\bibnamefont {Reichenthal}}, \bibinfo {author} {\bibfnamefont
			{H.}~\bibnamefont {Rodriguez}}, \bibinfo {author} {\bibfnamefont
			{J.}~\bibnamefont {Rouzaud}}, \bibinfo {author} {\bibfnamefont
			{A.}~\bibnamefont {Rucker}}, \bibinfo {author} {\bibfnamefont
			{Y.}~\bibnamefont {Saito}}, \bibinfo {author} {\bibfnamefont
			{M.}~\bibnamefont {Samara}}, \bibinfo {author} {\bibfnamefont {J.-A.}\
			\bibnamefont {Sauvaud}}, \bibinfo {author} {\bibfnamefont {D.}~\bibnamefont
			{Schuster}}, \bibinfo {author} {\bibfnamefont {M.}~\bibnamefont {Shappirio}},
		\bibinfo {author} {\bibfnamefont {K.}~\bibnamefont {Shelton}}, \bibinfo
		{author} {\bibfnamefont {D.}~\bibnamefont {Sher}}, \bibinfo {author}
		{\bibfnamefont {D.}~\bibnamefont {Smith}}, \bibinfo {author} {\bibfnamefont
			{K.}~\bibnamefont {Smith}}, \bibinfo {author} {\bibfnamefont
			{S.}~\bibnamefont {Smith}}, \bibinfo {author} {\bibfnamefont
			{D.}~\bibnamefont {Steinfeld}}, \bibinfo {author} {\bibfnamefont
			{R.}~\bibnamefont {Szymkiewicz}}, \bibinfo {author} {\bibfnamefont
			{K.}~\bibnamefont {Tanimoto}}, \bibinfo {author} {\bibfnamefont
			{J.}~\bibnamefont {Taylor}}, \bibinfo {author} {\bibfnamefont
			{C.}~\bibnamefont {Tucker}}, \bibinfo {author} {\bibfnamefont
			{K.}~\bibnamefont {Tull}}, \bibinfo {author} {\bibfnamefont {A.}~\bibnamefont
			{Uhl}}, \bibinfo {author} {\bibfnamefont {J.}~\bibnamefont {Vloet}}, \bibinfo
		{author} {\bibfnamefont {P.}~\bibnamefont {Walpole}}, \bibinfo {author}
		{\bibfnamefont {S.}~\bibnamefont {Weidner}}, \bibinfo {author} {\bibfnamefont
			{D.}~\bibnamefont {White}}, \bibinfo {author} {\bibfnamefont
			{G.}~\bibnamefont {Winkert}}, \bibinfo {author} {\bibfnamefont {P.-S.}\
			\bibnamefont {Yeh}}, \ and\ \bibinfo {author} {\bibfnamefont
			{M.}~\bibnamefont {Zeuch}},\ }\href {\doibase 10.1007/s11214-016-0245-4}
	{\bibfield  {journal} {\bibinfo  {journal} {Space Science Reviews}\ }\textbf
		{\bibinfo {volume} {199}},\ \bibinfo {pages} {331} (\bibinfo {year}
		{2016})}\BibitemShut {NoStop}%
	\bibitem [{\citenamefont {Russell}\ \emph {et~al.}(2016)\citenamefont
		{Russell}, \citenamefont {Anderson}, \citenamefont {Baumjohann},
		\citenamefont {Bromund}, \citenamefont {Dearborn}, \citenamefont {Fischer},
		\citenamefont {Le}, \citenamefont {Leinweber}, \citenamefont {Leneman},
		\citenamefont {Magnes}, \citenamefont {Means}, \citenamefont {Moldwin},
		\citenamefont {Nakamura}, \citenamefont {Pierce}, \citenamefont {Plaschke},
		\citenamefont {Rowe}, \citenamefont {Slavin}, \citenamefont {Strangeway},
		\citenamefont {Torbert}, \citenamefont {Hagen}, \citenamefont {Jernej},
		\citenamefont {Valavanoglou},\ and\ \citenamefont
		{Richter}}]{Russell2016SSR}%
	\BibitemOpen
	\bibfield  {author} {\bibinfo {author} {\bibfnamefont {C.~T.}\ \bibnamefont
			{Russell}}, \bibinfo {author} {\bibfnamefont {B.~J.}\ \bibnamefont
			{Anderson}}, \bibinfo {author} {\bibfnamefont {W.}~\bibnamefont
			{Baumjohann}}, \bibinfo {author} {\bibfnamefont {K.~R.}\ \bibnamefont
			{Bromund}}, \bibinfo {author} {\bibfnamefont {D.}~\bibnamefont {Dearborn}},
		\bibinfo {author} {\bibfnamefont {D.}~\bibnamefont {Fischer}}, \bibinfo
		{author} {\bibfnamefont {G.}~\bibnamefont {Le}}, \bibinfo {author}
		{\bibfnamefont {H.~K.}\ \bibnamefont {Leinweber}}, \bibinfo {author}
		{\bibfnamefont {D.}~\bibnamefont {Leneman}}, \bibinfo {author} {\bibfnamefont
			{W.}~\bibnamefont {Magnes}}, \bibinfo {author} {\bibfnamefont {J.~D.}\
			\bibnamefont {Means}}, \bibinfo {author} {\bibfnamefont {M.~B.}\ \bibnamefont
			{Moldwin}}, \bibinfo {author} {\bibfnamefont {R.}~\bibnamefont {Nakamura}},
		\bibinfo {author} {\bibfnamefont {D.}~\bibnamefont {Pierce}}, \bibinfo
		{author} {\bibfnamefont {F.}~\bibnamefont {Plaschke}}, \bibinfo {author}
		{\bibfnamefont {K.~M.}\ \bibnamefont {Rowe}}, \bibinfo {author}
		{\bibfnamefont {J.~A.}\ \bibnamefont {Slavin}}, \bibinfo {author}
		{\bibfnamefont {R.~J.}\ \bibnamefont {Strangeway}}, \bibinfo {author}
		{\bibfnamefont {R.}~\bibnamefont {Torbert}}, \bibinfo {author} {\bibfnamefont
			{C.}~\bibnamefont {Hagen}}, \bibinfo {author} {\bibfnamefont
			{I.}~\bibnamefont {Jernej}}, \bibinfo {author} {\bibfnamefont
			{A.}~\bibnamefont {Valavanoglou}}, \ and\ \bibinfo {author} {\bibfnamefont
			{I.}~\bibnamefont {Richter}},\ }\href {\doibase 10.1007/s11214-014-0057-3}
	{\bibfield  {journal} {\bibinfo  {journal} {Space Science Reviews}\ }\textbf
		{\bibinfo {volume} {199}},\ \bibinfo {pages} {189} (\bibinfo {year}
		{2016})}\BibitemShut {NoStop}%
	\bibitem [{\citenamefont {{Yang}}\ \emph
		{et~al.}(2017{\natexlab{a}})\citenamefont {{Yang}}, \citenamefont
		{{Matthaeus}}, \citenamefont {{Parashar}}, \citenamefont {{Wu}},
		\citenamefont {{Wan}}, \citenamefont {{Shi}}, \citenamefont {{Chen}},
		\citenamefont {{Roytershteyn}},\ and\ \citenamefont
		{{Daughton}}}]{Yang2017PRE}%
	\BibitemOpen
	\bibfield  {author} {\bibinfo {author} {\bibfnamefont {Y.}~\bibnamefont
			{{Yang}}}, \bibinfo {author} {\bibfnamefont {W.~H.}\ \bibnamefont
			{{Matthaeus}}}, \bibinfo {author} {\bibfnamefont {T.~N.}\ \bibnamefont
			{{Parashar}}}, \bibinfo {author} {\bibfnamefont {P.}~\bibnamefont {{Wu}}},
		\bibinfo {author} {\bibfnamefont {M.}~\bibnamefont {{Wan}}}, \bibinfo
		{author} {\bibfnamefont {Y.}~\bibnamefont {{Shi}}}, \bibinfo {author}
		{\bibfnamefont {S.}~\bibnamefont {{Chen}}}, \bibinfo {author} {\bibfnamefont
			{V.}~\bibnamefont {{Roytershteyn}}}, \ and\ \bibinfo {author} {\bibfnamefont
			{W.}~\bibnamefont {{Daughton}}},\ }\href {\doibase
		10.1103/PhysRevE.95.061201} {\bibfield  {journal} {\bibinfo  {journal} {Phys.
				Rev. E}\ }\textbf {\bibinfo {volume} {95}},\ \bibinfo {pages} {061201}
		(\bibinfo {year} {2017}{\natexlab{a}})}\BibitemShut {NoStop}%
	\bibitem [{\citenamefont {{Yang}}\ \emph
		{et~al.}(2017{\natexlab{b}})\citenamefont {{Yang}}, \citenamefont
		{{Matthaeus}}, \citenamefont {{Parashar}}, \citenamefont {{Haggerty}},
		\citenamefont {{Roytershteyn}}, \citenamefont {{Daughton}}, \citenamefont
		{{Wan}}, \citenamefont {{Shi}},\ and\ \citenamefont {{Chen}}}]{Yang2017PoP}%
	\BibitemOpen
	\bibfield  {author} {\bibinfo {author} {\bibfnamefont {Y.}~\bibnamefont
			{{Yang}}}, \bibinfo {author} {\bibfnamefont {W.~H.}\ \bibnamefont
			{{Matthaeus}}}, \bibinfo {author} {\bibfnamefont {T.~N.}\ \bibnamefont
			{{Parashar}}}, \bibinfo {author} {\bibfnamefont {C.~C.}\ \bibnamefont
			{{Haggerty}}}, \bibinfo {author} {\bibfnamefont {V.}~\bibnamefont
			{{Roytershteyn}}}, \bibinfo {author} {\bibfnamefont {W.}~\bibnamefont
			{{Daughton}}}, \bibinfo {author} {\bibfnamefont {M.}~\bibnamefont {{Wan}}},
		\bibinfo {author} {\bibfnamefont {Y.}~\bibnamefont {{Shi}}}, \ and\ \bibinfo
		{author} {\bibfnamefont {S.}~\bibnamefont {{Chen}}},\ }\href {\doibase
		10.1063/1.4990421} {\bibfield  {journal} {\bibinfo  {journal} {Phys.
				Plasmas}\ }\textbf {\bibinfo {volume} {24}},\ \bibinfo {pages} {072306}
		(\bibinfo {year} {2017}{\natexlab{b}})},\ \Eprint
	{http://arxiv.org/abs/1705.02054} {arXiv:1705.02054 [physics.plasm-ph]}
	\BibitemShut {NoStop}%
	\bibitem [{\citenamefont {Yang}\ \emph {et~al.}(2019)\citenamefont {Yang},
		\citenamefont {Wan}, \citenamefont {Matthaeus}, \citenamefont
		{Sorriso-Valvo}, \citenamefont {Parashar}, \citenamefont {Lu}, \citenamefont
		{Shi},\ and\ \citenamefont {Chen}}]{YangMNRAS18}%
	\BibitemOpen
	\bibfield  {author} {\bibinfo {author} {\bibfnamefont {Y.}~\bibnamefont
			{Yang}}, \bibinfo {author} {\bibfnamefont {M.}~\bibnamefont {Wan}}, \bibinfo
		{author} {\bibfnamefont {W.~H.}\ \bibnamefont {Matthaeus}}, \bibinfo {author}
		{\bibfnamefont {L.}~\bibnamefont {Sorriso-Valvo}}, \bibinfo {author}
		{\bibfnamefont {T.~N.}\ \bibnamefont {Parashar}}, \bibinfo {author}
		{\bibfnamefont {Q.}~\bibnamefont {Lu}}, \bibinfo {author} {\bibfnamefont
			{Y.}~\bibnamefont {Shi}}, \ and\ \bibinfo {author} {\bibfnamefont
			{S.}~\bibnamefont {Chen}},\ }\href@noop {} {\bibfield  {journal} {\bibinfo
			{journal} {Monthly Notices of the Royal Astronomical Society}\ }\textbf
		{\bibinfo {volume} {482}},\ \bibinfo {pages} {4933} (\bibinfo {year}
		{2019})}\BibitemShut {NoStop}%
	\bibitem [{\citenamefont {Del~Sarto}\ \emph {et~al.}(2016)\citenamefont
		{Del~Sarto}, \citenamefont {Pegoraro},\ and\ \citenamefont
		{Califano}}]{DelSartoPRE16}%
	\BibitemOpen
	\bibfield  {author} {\bibinfo {author} {\bibfnamefont {D.}~\bibnamefont
			{Del~Sarto}}, \bibinfo {author} {\bibfnamefont {F.}~\bibnamefont {Pegoraro}},
		\ and\ \bibinfo {author} {\bibfnamefont {F.}~\bibnamefont {Califano}},\
	}\href@noop {} {\bibfield  {journal} {\bibinfo  {journal} {Physical Review
			E}\ }\textbf {\bibinfo {volume} {93}},\ \bibinfo {pages} {053203} (\bibinfo
	{year} {2016})}\BibitemShut {NoStop}%
\bibitem [{\citenamefont {Del~Sarto}\ and\ \citenamefont
	{Pegoraro}(2017)}]{DelSartoMNRAS17}%
\BibitemOpen
\bibfield  {author} {\bibinfo {author} {\bibfnamefont {D.}~\bibnamefont
		{Del~Sarto}}\ and\ \bibinfo {author} {\bibfnamefont {F.}~\bibnamefont
		{Pegoraro}},\ }\href@noop {} {\bibfield  {journal} {\bibinfo  {journal}
		{Monthly Notices of the Royal Astronomical Society}\ }\textbf {\bibinfo
		{volume} {475}},\ \bibinfo {pages} {181} (\bibinfo {year}
	{2017})}\BibitemShut {NoStop}%
\bibitem [{\citenamefont {Sitnov}\ \emph {et~al.}(2018)\citenamefont {Sitnov},
	\citenamefont {Merkin}, \citenamefont {Roytershteyn},\ and\ \citenamefont
	{Swisdak}}]{SitnovGRL18}%
\BibitemOpen
\bibfield  {author} {\bibinfo {author} {\bibfnamefont {M.}~\bibnamefont
		{Sitnov}}, \bibinfo {author} {\bibfnamefont {V.}~\bibnamefont {Merkin}},
	\bibinfo {author} {\bibfnamefont {V.}~\bibnamefont {Roytershteyn}}, \ and\
	\bibinfo {author} {\bibfnamefont {M.}~\bibnamefont {Swisdak}},\ }\href@noop
{} {\bibfield  {journal} {\bibinfo  {journal} {Geophysical Research Letters}\
	}\textbf {\bibinfo {volume} {45}},\ \bibinfo {pages} {4639} (\bibinfo {year}
	{2018})}\BibitemShut {NoStop}%
\bibitem [{\citenamefont {Pezzi}\ \emph {et~al.}(2019)\citenamefont {Pezzi},
	\citenamefont {Yang}, \citenamefont {Valentini}, \citenamefont {Servidio},
	\citenamefont {Chasapis}, \citenamefont {Matthaeus},\ and\ \citenamefont
	{Veltri}}]{PezziApJ19}%
\BibitemOpen
\bibfield  {author} {\bibinfo {author} {\bibfnamefont {O.}~\bibnamefont
		{Pezzi}}, \bibinfo {author} {\bibfnamefont {Y.}~\bibnamefont {Yang}},
	\bibinfo {author} {\bibfnamefont {F.}~\bibnamefont {Valentini}}, \bibinfo
	{author} {\bibfnamefont {S.}~\bibnamefont {Servidio}}, \bibinfo {author}
	{\bibfnamefont {A.}~\bibnamefont {Chasapis}}, \bibinfo {author}
	{\bibfnamefont {W.}~\bibnamefont {Matthaeus}}, \ and\ \bibinfo {author}
	{\bibfnamefont {P.}~\bibnamefont {Veltri}},\ }\href@noop {} {\bibfield
	{journal} {\bibinfo  {journal} {arXiv preprint arXiv:1904.07715}\ } (\bibinfo
	{year} {2019})}\BibitemShut {NoStop}%
\bibitem [{\citenamefont {{Del Sarto}}\ \emph {et~al.}(2016)\citenamefont {{Del
			Sarto}}, \citenamefont {{Pegoraro}},\ and\ \citenamefont
	{{Califano}}}]{DelSartoPegoraro16}%
\BibitemOpen
\bibfield  {author} {\bibinfo {author} {\bibfnamefont {D.}~\bibnamefont {{Del
				Sarto}}}, \bibinfo {author} {\bibfnamefont {F.}~\bibnamefont {{Pegoraro}}}, \
	and\ \bibinfo {author} {\bibfnamefont {F.}~\bibnamefont {{Califano}}},\
}\href {\doibase 10.1103/PhysRevE.93.053203} {\bibfield  {journal} {\bibinfo
	{journal} {\pre}\ }\textbf {\bibinfo {volume} {93}},\ \bibinfo {eid} {053203}
(\bibinfo {year} {2016})},\ \Eprint {http://arxiv.org/abs/1507.04895}
{arXiv:1507.04895 [physics.plasm-ph]} \BibitemShut {NoStop}%
\bibitem [{\citenamefont {Matthaeus}(1982)}]{MatthaeusGRL82}%
\BibitemOpen
\bibfield  {author} {\bibinfo {author} {\bibfnamefont {W.~H.}\ \bibnamefont
		{Matthaeus}},\ }\href@noop {} {\bibfield  {journal} {\bibinfo  {journal}
		{Geophysical Research Letters}\ }\textbf {\bibinfo {volume} {9}},\ \bibinfo
	{pages} {660} (\bibinfo {year} {1982})}\BibitemShut {NoStop}%
\bibitem [{\citenamefont {Parashar}\ and\ \citenamefont
	{Matthaeus}(2016)}]{Parashar2016ApJ}%
\BibitemOpen
\bibfield  {author} {\bibinfo {author} {\bibfnamefont {T.~N.}\ \bibnamefont
		{Parashar}}\ and\ \bibinfo {author} {\bibfnamefont {W.~H.}\ \bibnamefont
		{Matthaeus}},\ }\href {\doibase 10.3847/0004-637X/832/1/57} {\bibfield
	{journal} {\bibinfo  {journal} {The Astrophysical Journal}\ }\textbf
	{\bibinfo {volume} {832}},\ \bibinfo {pages} {57} (\bibinfo {year}
	{2016})}\BibitemShut {NoStop}%
\bibitem [{\citenamefont {{Zhdankin}}\ \emph {et~al.}(2013)\citenamefont
	{{Zhdankin}}, \citenamefont {{Uzdensky}}, \citenamefont {{Perez}},\ and\
	\citenamefont {{Boldyrev}}}]{Zhdankin2013ApJ}%
\BibitemOpen
\bibfield  {author} {\bibinfo {author} {\bibfnamefont {V.}~\bibnamefont
		{{Zhdankin}}}, \bibinfo {author} {\bibfnamefont {D.~A.}\ \bibnamefont
		{{Uzdensky}}}, \bibinfo {author} {\bibfnamefont {J.~C.}\ \bibnamefont
		{{Perez}}}, \ and\ \bibinfo {author} {\bibfnamefont {S.}~\bibnamefont
		{{Boldyrev}}},\ }\href {\doibase 10.1088/0004-637X/771/2/124} {\bibfield
	{journal} {\bibinfo  {journal} {The Astrophysical Journal}\ }\textbf
	{\bibinfo {volume} {771}},\ \bibinfo {eid} {124} (\bibinfo {year}
	{2013})}\BibitemShut {NoStop}%
\bibitem [{\citenamefont {{Wan}}\ \emph {et~al.}(2014)\citenamefont {{Wan}},
	\citenamefont {{Rappazzo}}, \citenamefont {{Matthaeus}}, \citenamefont
	{{Servidio}},\ and\ \citenamefont {{Oughton}}}]{Wan2014ApJ}%
\BibitemOpen
\bibfield  {author} {\bibinfo {author} {\bibfnamefont {M.}~\bibnamefont
		{{Wan}}}, \bibinfo {author} {\bibfnamefont {A.~F.}\ \bibnamefont
		{{Rappazzo}}}, \bibinfo {author} {\bibfnamefont {W.~H.}\ \bibnamefont
		{{Matthaeus}}}, \bibinfo {author} {\bibfnamefont {S.}~\bibnamefont
		{{Servidio}}}, \ and\ \bibinfo {author} {\bibfnamefont {S.}~\bibnamefont
		{{Oughton}}},\ }\href {\doibase 10.1088/0004-637X/797/1/63} {\bibfield
	{journal} {\bibinfo  {journal} {The Astrophysical Journal}\ }\textbf
	{\bibinfo {volume} {797}},\ \bibinfo {eid} {63} (\bibinfo {year}
	{2014})}\BibitemShut {NoStop}%
\bibitem [{\citenamefont {Chasapis}\ \emph
	{et~al.}(2018{\natexlab{a}})\citenamefont {Chasapis}, \citenamefont {Yang},
	\citenamefont {Matthaeus}, \citenamefont {Parashar}, \citenamefont
	{Haggerty}, \citenamefont {Burch}, \citenamefont {Moore}, \citenamefont
	{Pollock}, \citenamefont {Dorelli}, \citenamefont {Gershman}, \citenamefont
	{Torbert},\ and\ \citenamefont {Russell}}]{ChasapisApJ18}%
\BibitemOpen
\bibfield  {author} {\bibinfo {author} {\bibfnamefont {A.}~\bibnamefont
		{Chasapis}}, \bibinfo {author} {\bibfnamefont {Y.}~\bibnamefont {Yang}},
	\bibinfo {author} {\bibfnamefont {W.~H.}\ \bibnamefont {Matthaeus}}, \bibinfo
	{author} {\bibfnamefont {T.~N.}\ \bibnamefont {Parashar}}, \bibinfo {author}
	{\bibfnamefont {C.~C.}\ \bibnamefont {Haggerty}}, \bibinfo {author}
	{\bibfnamefont {J.~L.}\ \bibnamefont {Burch}}, \bibinfo {author}
	{\bibfnamefont {T.~E.}\ \bibnamefont {Moore}}, \bibinfo {author}
	{\bibfnamefont {C.~J.}\ \bibnamefont {Pollock}}, \bibinfo {author}
	{\bibfnamefont {J.}~\bibnamefont {Dorelli}}, \bibinfo {author} {\bibfnamefont
		{D.~J.}\ \bibnamefont {Gershman}}, \bibinfo {author} {\bibfnamefont {R.~B.}\
		\bibnamefont {Torbert}}, \ and\ \bibinfo {author} {\bibfnamefont {C.~T.}\
		\bibnamefont {Russell}},\ }\href
{http://stacks.iop.org/0004-637X/862/i=1/a=32} {\bibfield  {journal}
	{\bibinfo  {journal} {The Astrophysical Journal}\ }\textbf {\bibinfo {volume}
		{862}},\ \bibinfo {pages} {32} (\bibinfo {year}
	{2018}{\natexlab{a}})}\BibitemShut {NoStop}%
\bibitem [{\citenamefont {Franci}\ \emph {et~al.}(2016)\citenamefont {Franci},
	\citenamefont {Hellinger}, \citenamefont {Matteini}, \citenamefont
	{Verdini},\ and\ \citenamefont {Landi}}]{FranciSW14}%
\BibitemOpen
\bibfield  {author} {\bibinfo {author} {\bibfnamefont {L.}~\bibnamefont
		{Franci}}, \bibinfo {author} {\bibfnamefont {P.}~\bibnamefont {Hellinger}},
	\bibinfo {author} {\bibfnamefont {L.}~\bibnamefont {Matteini}}, \bibinfo
	{author} {\bibfnamefont {A.}~\bibnamefont {Verdini}}, \ and\ \bibinfo
	{author} {\bibfnamefont {S.}~\bibnamefont {Landi}},\ }in\ \href@noop {}
{\emph {\bibinfo {booktitle} {SOLAR WIND 14: Proceedings of the Fourteenth
			International Solar Wind Conference}}},\ Vol.\ \bibinfo {volume} {1720}\
(\bibinfo {organization} {AIP Publishing},\ \bibinfo {year} {2016})\ p.\
\bibinfo {pages} {040003}\BibitemShut {NoStop}%
\bibitem [{\citenamefont {{Huang}}\ \emph
	{et~al.}(2017{\natexlab{a}})\citenamefont {{Huang}}, \citenamefont
	{{Sahraoui}}, \citenamefont {{Yuan}}, \citenamefont {{He}}, \citenamefont
	{{Zhao}}, \citenamefont {{Le Contel}}, \citenamefont {{Deng}}, \citenamefont
	{{Zhou}}, \citenamefont {{Fu}}, \citenamefont {{Shi}}, \citenamefont
	{{Lavraud}}, \citenamefont {{Pang}}, \citenamefont {{Yang}}, \citenamefont
	{{Wang}}, \citenamefont {{Li}}, \citenamefont {{Yu}}, \citenamefont
	{{Pollock}}, \citenamefont {{Giles}}, \citenamefont {{Torbert}},
	\citenamefont {{Russell}}, \citenamefont {{Goodrich}}, \citenamefont
	{{Gershman}}, \citenamefont {{Moore}}, \citenamefont {{Ergun}}, \citenamefont
	{{Khotyaintsev}}, \citenamefont {{Lindqvist}}, \citenamefont {{Strangeway}},
	\citenamefont {{Magnes}}, \citenamefont {{Bromund}}, \citenamefont
	{{Leinweber}}, \citenamefont {{Plaschke}}, \citenamefont {{Anderson}},\ and\
	\citenamefont {{Burch}}}]{Huang2017bApJL}%
\BibitemOpen
\bibfield  {author} {\bibinfo {author} {\bibfnamefont {S.~Y.}\ \bibnamefont
		{{Huang}}}, \bibinfo {author} {\bibfnamefont {F.}~\bibnamefont {{Sahraoui}}},
	\bibinfo {author} {\bibfnamefont {Z.~G.}\ \bibnamefont {{Yuan}}}, \bibinfo
	{author} {\bibfnamefont {J.~S.}\ \bibnamefont {{He}}}, \bibinfo {author}
	{\bibfnamefont {J.~S.}\ \bibnamefont {{Zhao}}}, \bibinfo {author}
	{\bibfnamefont {O.}~\bibnamefont {{Le Contel}}}, \bibinfo {author}
	{\bibfnamefont {X.~H.}\ \bibnamefont {{Deng}}}, \bibinfo {author}
	{\bibfnamefont {M.}~\bibnamefont {{Zhou}}}, \bibinfo {author} {\bibfnamefont
		{H.~S.}\ \bibnamefont {{Fu}}}, \bibinfo {author} {\bibfnamefont {Q.~Q.}\
		\bibnamefont {{Shi}}}, \bibinfo {author} {\bibfnamefont {B.}~\bibnamefont
		{{Lavraud}}}, \bibinfo {author} {\bibfnamefont {Y.}~\bibnamefont {{Pang}}},
	\bibinfo {author} {\bibfnamefont {J.}~\bibnamefont {{Yang}}}, \bibinfo
	{author} {\bibfnamefont {D.~D.}\ \bibnamefont {{Wang}}}, \bibinfo {author}
	{\bibfnamefont {H.~M.}\ \bibnamefont {{Li}}}, \bibinfo {author}
	{\bibfnamefont {X.~D.}\ \bibnamefont {{Yu}}}, \bibinfo {author}
	{\bibfnamefont {C.~J.}\ \bibnamefont {{Pollock}}}, \bibinfo {author}
	{\bibfnamefont {B.~L.}\ \bibnamefont {{Giles}}}, \bibinfo {author}
	{\bibfnamefont {R.~B.}\ \bibnamefont {{Torbert}}}, \bibinfo {author}
	{\bibfnamefont {C.~T.}\ \bibnamefont {{Russell}}}, \bibinfo {author}
	{\bibfnamefont {K.~A.}\ \bibnamefont {{Goodrich}}}, \bibinfo {author}
	{\bibfnamefont {D.~J.}\ \bibnamefont {{Gershman}}}, \bibinfo {author}
	{\bibfnamefont {T.~E.}\ \bibnamefont {{Moore}}}, \bibinfo {author}
	{\bibfnamefont {R.~E.}\ \bibnamefont {{Ergun}}}, \bibinfo {author}
	{\bibfnamefont {Y.~V.}\ \bibnamefont {{Khotyaintsev}}}, \bibinfo {author}
	{\bibfnamefont {P.~A.}\ \bibnamefont {{Lindqvist}}}, \bibinfo {author}
	{\bibfnamefont {R.~J.}\ \bibnamefont {{Strangeway}}}, \bibinfo {author}
	{\bibfnamefont {W.}~\bibnamefont {{Magnes}}}, \bibinfo {author}
	{\bibfnamefont {K.}~\bibnamefont {{Bromund}}}, \bibinfo {author}
	{\bibfnamefont {H.}~\bibnamefont {{Leinweber}}}, \bibinfo {author}
	{\bibfnamefont {F.}~\bibnamefont {{Plaschke}}}, \bibinfo {author}
	{\bibfnamefont {B.~J.}\ \bibnamefont {{Anderson}}}, \ and\ \bibinfo {author}
	{\bibfnamefont {J.~L.}\ \bibnamefont {{Burch}}},\ }\href {\doibase
	10.3847/2041-8213/aa5f50} {\bibfield  {journal} {\bibinfo  {journal}
		{Astrophys. J. Lett.}\ }\textbf {\bibinfo {volume} {836}},\ \bibinfo {eid}
	{L27} (\bibinfo {year} {2017}{\natexlab{a}})}\BibitemShut {NoStop}%
\bibitem [{\citenamefont {{Huang}}\ \emph
	{et~al.}(2017{\natexlab{b}})\citenamefont {{Huang}}, \citenamefont {{Du}},
	\citenamefont {{Sahraoui}}, \citenamefont {{Yuan}}, \citenamefont {{He}},
	\citenamefont {{Zhao}}, \citenamefont {{Le Contel}}, \citenamefont
	{{Breuillard}}, \citenamefont {{Wang}}, \citenamefont {{Yu}}, \citenamefont
	{{Deng}}, \citenamefont {{Fu}}, \citenamefont {{Zhou}}, \citenamefont
	{{Pollock}}, \citenamefont {{Torbert}}, \citenamefont {{Russell}},\ and\
	\citenamefont {{Burch}}}]{Huang2017JGR}%
\BibitemOpen
\bibfield  {author} {\bibinfo {author} {\bibfnamefont {S.~Y.}\ \bibnamefont
		{{Huang}}}, \bibinfo {author} {\bibfnamefont {J.~W.}\ \bibnamefont {{Du}}},
	\bibinfo {author} {\bibfnamefont {F.}~\bibnamefont {{Sahraoui}}}, \bibinfo
	{author} {\bibfnamefont {Z.~G.}\ \bibnamefont {{Yuan}}}, \bibinfo {author}
	{\bibfnamefont {J.~S.}\ \bibnamefont {{He}}}, \bibinfo {author}
	{\bibfnamefont {J.~S.}\ \bibnamefont {{Zhao}}}, \bibinfo {author}
	{\bibfnamefont {O.}~\bibnamefont {{Le Contel}}}, \bibinfo {author}
	{\bibfnamefont {H.}~\bibnamefont {{Breuillard}}}, \bibinfo {author}
	{\bibfnamefont {D.~D.}\ \bibnamefont {{Wang}}}, \bibinfo {author}
	{\bibfnamefont {X.~D.}\ \bibnamefont {{Yu}}}, \bibinfo {author}
	{\bibfnamefont {X.~H.}\ \bibnamefont {{Deng}}}, \bibinfo {author}
	{\bibfnamefont {H.~S.}\ \bibnamefont {{Fu}}}, \bibinfo {author}
	{\bibfnamefont {M.}~\bibnamefont {{Zhou}}}, \bibinfo {author} {\bibfnamefont
		{C.~J.}\ \bibnamefont {{Pollock}}}, \bibinfo {author} {\bibfnamefont {R.~B.}\
		\bibnamefont {{Torbert}}}, \bibinfo {author} {\bibfnamefont {C.~T.}\
		\bibnamefont {{Russell}}}, \ and\ \bibinfo {author} {\bibfnamefont {J.~L.}\
		\bibnamefont {{Burch}}},\ }\href {\doibase 10.1002/2017JA024415} {\bibfield
	{journal} {\bibinfo  {journal} {Journal of Geophysical Research (Space
			Physics)}\ }\textbf {\bibinfo {volume} {122}},\ \bibinfo {pages} {8577}
	(\bibinfo {year} {2017}{\natexlab{b}})}\BibitemShut {NoStop}%
\bibitem [{\citenamefont {{Huang}}\ \emph {et~al.}(2018)\citenamefont
	{{Huang}}, \citenamefont {{Sahraoui}}, \citenamefont {{Yuan}}, \citenamefont
	{{Le Contel}}, \citenamefont {{Breuillard}}, \citenamefont {{He}},
	\citenamefont {{Zhao}}, \citenamefont {{Fu}}, \citenamefont {{Zhou}},
	\citenamefont {{Deng}}, \citenamefont {{Wang}}, \citenamefont {{Du}},
	\citenamefont {{Yu}}, \citenamefont {{Wang}}, \citenamefont {{Pollock}},
	\citenamefont {{Torbert}},\ and\ \citenamefont {{Burch}}}]{Huang2018ApJ}%
\BibitemOpen
\bibfield  {author} {\bibinfo {author} {\bibfnamefont {S.~Y.}\ \bibnamefont
		{{Huang}}}, \bibinfo {author} {\bibfnamefont {F.}~\bibnamefont {{Sahraoui}}},
	\bibinfo {author} {\bibfnamefont {Z.~G.}\ \bibnamefont {{Yuan}}}, \bibinfo
	{author} {\bibfnamefont {O.}~\bibnamefont {{Le Contel}}}, \bibinfo {author}
	{\bibfnamefont {H.}~\bibnamefont {{Breuillard}}}, \bibinfo {author}
	{\bibfnamefont {J.~S.}\ \bibnamefont {{He}}}, \bibinfo {author}
	{\bibfnamefont {J.~S.}\ \bibnamefont {{Zhao}}}, \bibinfo {author}
	{\bibfnamefont {H.~S.}\ \bibnamefont {{Fu}}}, \bibinfo {author}
	{\bibfnamefont {M.}~\bibnamefont {{Zhou}}}, \bibinfo {author} {\bibfnamefont
		{X.~H.}\ \bibnamefont {{Deng}}}, \bibinfo {author} {\bibfnamefont {X.~Y.}\
		\bibnamefont {{Wang}}}, \bibinfo {author} {\bibfnamefont {J.~W.}\
		\bibnamefont {{Du}}}, \bibinfo {author} {\bibfnamefont {X.~D.}\ \bibnamefont
		{{Yu}}}, \bibinfo {author} {\bibfnamefont {D.~D.}\ \bibnamefont {{Wang}}},
	\bibinfo {author} {\bibfnamefont {C.~J.}\ \bibnamefont {{Pollock}}}, \bibinfo
	{author} {\bibfnamefont {R.~B.}\ \bibnamefont {{Torbert}}}, \ and\ \bibinfo
	{author} {\bibfnamefont {J.~L.}\ \bibnamefont {{Burch}}},\ }\href {\doibase
	10.3847/1538-4357/aac831} {\bibfield  {journal} {\bibinfo  {journal} {The
			Astrophysical Journal}\ }\textbf {\bibinfo {volume} {861}},\ \bibinfo {eid}
	{29} (\bibinfo {year} {2018})}\BibitemShut {NoStop}%
\bibitem [{\citenamefont {Shue}\ \emph {et~al.}(1998)\citenamefont {Shue},
	\citenamefont {Song}, \citenamefont {Russell}, \citenamefont {Steinberg},
	\citenamefont {Chao}, \citenamefont {Zastenker}, \citenamefont {Vaisberg},
	\citenamefont {Kokubun}, \citenamefont {Singer}, \citenamefont {Detman},\
	and\ \citenamefont {Kawano}}]{Shue1998JGR}%
\BibitemOpen
\bibfield  {author} {\bibinfo {author} {\bibfnamefont {J.-H.}\ \bibnamefont
		{Shue}}, \bibinfo {author} {\bibfnamefont {P.}~\bibnamefont {Song}}, \bibinfo
	{author} {\bibfnamefont {C.~T.}\ \bibnamefont {Russell}}, \bibinfo {author}
	{\bibfnamefont {J.~T.}\ \bibnamefont {Steinberg}}, \bibinfo {author}
	{\bibfnamefont {J.~K.}\ \bibnamefont {Chao}}, \bibinfo {author}
	{\bibfnamefont {G.}~\bibnamefont {Zastenker}}, \bibinfo {author}
	{\bibfnamefont {O.~L.}\ \bibnamefont {Vaisberg}}, \bibinfo {author}
	{\bibfnamefont {S.}~\bibnamefont {Kokubun}}, \bibinfo {author} {\bibfnamefont
		{H.~J.}\ \bibnamefont {Singer}}, \bibinfo {author} {\bibfnamefont {T.~R.}\
		\bibnamefont {Detman}}, \ and\ \bibinfo {author} {\bibfnamefont
		{H.}~\bibnamefont {Kawano}},\ }\href {\doibase 10.1029/98JA01103} {\bibfield
	{journal} {\bibinfo  {journal} {Journal of Geophysical Research: Space
			Physics}\ }\textbf {\bibinfo {volume} {103}},\ \bibinfo {pages} {17691}
	(\bibinfo {year} {1998})}\BibitemShut {NoStop}%
\bibitem [{\citenamefont {Farris}\ and\ \citenamefont
	{Russell}(1994)}]{Farris1994JGR}%
\BibitemOpen
\bibfield  {author} {\bibinfo {author} {\bibfnamefont {M.~H.}\ \bibnamefont
		{Farris}}\ and\ \bibinfo {author} {\bibfnamefont {C.~T.}\ \bibnamefont
		{Russell}},\ }\href {\doibase 10.1029/94JA01020} {\bibfield  {journal}
	{\bibinfo  {journal} {Journal of Geophysical Research: Space Physics}\
	}\textbf {\bibinfo {volume} {99}},\ \bibinfo {pages} {17681} (\bibinfo {year}
	{1994})}\BibitemShut {NoStop}%
\bibitem [{\citenamefont {Parashar}\ \emph {et~al.}(2018)\citenamefont
	{Parashar}, \citenamefont {Chasapis}, \citenamefont {Bandyopadhyay},
	\citenamefont {Chhiber}, \citenamefont {Matthaeus}, \citenamefont {Maruca},
	\citenamefont {Shay}, \citenamefont {Burch}, \citenamefont {Moore},
	\citenamefont {Giles}, \citenamefont {Gershman}, \citenamefont {Pollock},
	\citenamefont {Torbert}, \citenamefont {Russell}, \citenamefont
	{Strangeway},\ and\ \citenamefont {Roytershteyn}}]{ParasharPRL18}%
\BibitemOpen
\bibfield  {author} {\bibinfo {author} {\bibfnamefont {T.~N.}\ \bibnamefont
		{Parashar}}, \bibinfo {author} {\bibfnamefont {A.}~\bibnamefont {Chasapis}},
	\bibinfo {author} {\bibfnamefont {R.}~\bibnamefont {Bandyopadhyay}}, \bibinfo
	{author} {\bibfnamefont {R.}~\bibnamefont {Chhiber}}, \bibinfo {author}
	{\bibfnamefont {W.~H.}\ \bibnamefont {Matthaeus}}, \bibinfo {author}
	{\bibfnamefont {B.}~\bibnamefont {Maruca}}, \bibinfo {author} {\bibfnamefont
		{M.~A.}\ \bibnamefont {Shay}}, \bibinfo {author} {\bibfnamefont {J.~L.}\
		\bibnamefont {Burch}}, \bibinfo {author} {\bibfnamefont {T.~E.}\ \bibnamefont
		{Moore}}, \bibinfo {author} {\bibfnamefont {B.~L.}\ \bibnamefont {Giles}},
	\bibinfo {author} {\bibfnamefont {D.~J.}\ \bibnamefont {Gershman}}, \bibinfo
	{author} {\bibfnamefont {C.~J.}\ \bibnamefont {Pollock}}, \bibinfo {author}
	{\bibfnamefont {R.~B.}\ \bibnamefont {Torbert}}, \bibinfo {author}
	{\bibfnamefont {C.~T.}\ \bibnamefont {Russell}}, \bibinfo {author}
	{\bibfnamefont {R.~J.}\ \bibnamefont {Strangeway}}, \ and\ \bibinfo {author}
	{\bibfnamefont {V.}~\bibnamefont {Roytershteyn}},\ }\href {\doibase
	10.1103/PhysRevLett.121.265101} {\bibfield  {journal} {\bibinfo  {journal}
		{Phys. Rev. Lett.}\ }\textbf {\bibinfo {volume} {121}},\ \bibinfo {pages}
	{265101} (\bibinfo {year} {2018})}\BibitemShut {NoStop}%
\bibitem [{\citenamefont {Huang}\ \emph {et~al.}(2017)\citenamefont {Huang},
	\citenamefont {Hadid}, \citenamefont {Sahraoui}, \citenamefont {Yuan},\ and\
	\citenamefont {Deng}}]{Huang2017ApJL}%
\BibitemOpen
\bibfield  {author} {\bibinfo {author} {\bibfnamefont {S.~Y.}\ \bibnamefont
		{Huang}}, \bibinfo {author} {\bibfnamefont {L.~Z.}\ \bibnamefont {Hadid}},
	\bibinfo {author} {\bibfnamefont {F.}~\bibnamefont {Sahraoui}}, \bibinfo
	{author} {\bibfnamefont {Z.~G.}\ \bibnamefont {Yuan}}, \ and\ \bibinfo
	{author} {\bibfnamefont {X.~H.}\ \bibnamefont {Deng}},\ }\href
{http://stacks.iop.org/2041-8205/836/i=1/a=L10} {\bibfield  {journal}
	{\bibinfo  {journal} {Astrophys. J. Lett.}\ }\textbf {\bibinfo {volume}
		{836}},\ \bibinfo {pages} {L10} (\bibinfo {year} {2017})}\BibitemShut
{NoStop}%
\bibitem [{\citenamefont {Dunlop}\ \emph {et~al.}(1988)\citenamefont {Dunlop},
	\citenamefont {Southwood}, \citenamefont {Glassmeier},\ and\ \citenamefont
	{Neubauer}}]{Dunlop1988ASR}%
\BibitemOpen
\bibfield  {author} {\bibinfo {author} {\bibfnamefont {M.}~\bibnamefont
		{Dunlop}}, \bibinfo {author} {\bibfnamefont {D.}~\bibnamefont {Southwood}},
	\bibinfo {author} {\bibfnamefont {K.-H.}\ \bibnamefont {Glassmeier}}, \ and\
	\bibinfo {author} {\bibfnamefont {F.}~\bibnamefont {Neubauer}},\ }\href
{\doibase https://doi.org/10.1016/0273-1177(88)90141-X} {\bibfield  {journal}
	{\bibinfo  {journal} {Advances in Space Research}\ }\textbf {\bibinfo
		{volume} {8}},\ \bibinfo {pages} {273} (\bibinfo {year} {1988})}\BibitemShut
{NoStop}%
\bibitem [{\citenamefont {Graham}\ \emph {et~al.}(2016)\citenamefont {Graham},
	\citenamefont {Khotyaintsev}, \citenamefont {Norgren}, \citenamefont
	{Vaivads}, \citenamefont {André}, \citenamefont {Lindqvist}, \citenamefont
	{Marklund}, \citenamefont {Ergun}, \citenamefont {Paterson}, \citenamefont
	{Gershman}, \citenamefont {Giles}, \citenamefont {Pollock}, \citenamefont
	{Dorelli}, \citenamefont {Avanov}, \citenamefont {Lavraud}, \citenamefont
	{Saito}, \citenamefont {Magnes}, \citenamefont {Russell}, \citenamefont
	{Strangeway}, \citenamefont {Torbert},\ and\ \citenamefont
	{Burch}}]{Graham2016GRL}%
\BibitemOpen
\bibfield  {author} {\bibinfo {author} {\bibfnamefont {D.~B.}\ \bibnamefont
		{Graham}}, \bibinfo {author} {\bibfnamefont {Y.~V.}\ \bibnamefont
		{Khotyaintsev}}, \bibinfo {author} {\bibfnamefont {C.}~\bibnamefont
		{Norgren}}, \bibinfo {author} {\bibfnamefont {A.}~\bibnamefont {Vaivads}},
	\bibinfo {author} {\bibfnamefont {M.}~\bibnamefont {André}}, \bibinfo
	{author} {\bibfnamefont {P.-A.}\ \bibnamefont {Lindqvist}}, \bibinfo {author}
	{\bibfnamefont {G.~T.}\ \bibnamefont {Marklund}}, \bibinfo {author}
	{\bibfnamefont {R.~E.}\ \bibnamefont {Ergun}}, \bibinfo {author}
	{\bibfnamefont {W.~R.}\ \bibnamefont {Paterson}}, \bibinfo {author}
	{\bibfnamefont {D.~J.}\ \bibnamefont {Gershman}}, \bibinfo {author}
	{\bibfnamefont {B.~L.}\ \bibnamefont {Giles}}, \bibinfo {author}
	{\bibfnamefont {C.~J.}\ \bibnamefont {Pollock}}, \bibinfo {author}
	{\bibfnamefont {J.~C.}\ \bibnamefont {Dorelli}}, \bibinfo {author}
	{\bibfnamefont {L.~A.}\ \bibnamefont {Avanov}}, \bibinfo {author}
	{\bibfnamefont {B.}~\bibnamefont {Lavraud}}, \bibinfo {author} {\bibfnamefont
		{Y.}~\bibnamefont {Saito}}, \bibinfo {author} {\bibfnamefont
		{W.}~\bibnamefont {Magnes}}, \bibinfo {author} {\bibfnamefont {C.~T.}\
		\bibnamefont {Russell}}, \bibinfo {author} {\bibfnamefont {R.~J.}\
		\bibnamefont {Strangeway}}, \bibinfo {author} {\bibfnamefont {R.~B.}\
		\bibnamefont {Torbert}}, \ and\ \bibinfo {author} {\bibfnamefont {J.~L.}\
		\bibnamefont {Burch}},\ }\href {\doibase 10.1002/2016GL068613} {\bibfield
	{journal} {\bibinfo  {journal} {Geophysical Research Letters}\ }\textbf
	{\bibinfo {volume} {43}},\ \bibinfo {pages} {4691} (\bibinfo {year}
	{2016})}\BibitemShut {NoStop}%
\bibitem [{\citenamefont {Gershman}\ \emph {et~al.}(2018)\citenamefont
	{Gershman}, \citenamefont {Vinas}, \citenamefont {Dorelli}, \citenamefont
	{Goldstein}, \citenamefont {Shuster}, \citenamefont {Avanov}, \citenamefont
	{Boardsen}, \citenamefont {Stawarz}, \citenamefont {Schwartz}, \citenamefont
	{Schiff}, \citenamefont {Lavraud}, \citenamefont {Saito}, \citenamefont
	{Paterson}, \citenamefont {Giles}, \citenamefont {Pollock}, \citenamefont
	{Strangeway}, \citenamefont {Russell}, \citenamefont {Torbert}, \citenamefont
	{Moore},\ and\ \citenamefont {Burch}}]{Gershman2018PoP}%
\BibitemOpen
\bibfield  {author} {\bibinfo {author} {\bibfnamefont {D.~J.}\ \bibnamefont
		{Gershman}}, \bibinfo {author} {\bibfnamefont {A.~F.}\ \bibnamefont {Vinas}},
	\bibinfo {author} {\bibfnamefont {J.~C.}\ \bibnamefont {Dorelli}}, \bibinfo
	{author} {\bibfnamefont {M.~L.}\ \bibnamefont {Goldstein}}, \bibinfo {author}
	{\bibfnamefont {J.}~\bibnamefont {Shuster}}, \bibinfo {author} {\bibfnamefont
		{L.~A.}\ \bibnamefont {Avanov}}, \bibinfo {author} {\bibfnamefont {S.~A.}\
		\bibnamefont {Boardsen}}, \bibinfo {author} {\bibfnamefont {J.~E.}\
		\bibnamefont {Stawarz}}, \bibinfo {author} {\bibfnamefont {S.~J.}\
		\bibnamefont {Schwartz}}, \bibinfo {author} {\bibfnamefont {C.}~\bibnamefont
		{Schiff}}, \bibinfo {author} {\bibfnamefont {B.}~\bibnamefont {Lavraud}},
	\bibinfo {author} {\bibfnamefont {Y.}~\bibnamefont {Saito}}, \bibinfo
	{author} {\bibfnamefont {W.~R.}\ \bibnamefont {Paterson}}, \bibinfo {author}
	{\bibfnamefont {B.~L.}\ \bibnamefont {Giles}}, \bibinfo {author}
	{\bibfnamefont {C.~J.}\ \bibnamefont {Pollock}}, \bibinfo {author}
	{\bibfnamefont {R.~J.}\ \bibnamefont {Strangeway}}, \bibinfo {author}
	{\bibfnamefont {C.~T.}\ \bibnamefont {Russell}}, \bibinfo {author}
	{\bibfnamefont {R.~B.}\ \bibnamefont {Torbert}}, \bibinfo {author}
	{\bibfnamefont {T.~E.}\ \bibnamefont {Moore}}, \ and\ \bibinfo {author}
	{\bibfnamefont {J.~L.}\ \bibnamefont {Burch}},\ }\href {\doibase
	10.1063/1.5009158} {\bibfield  {journal} {\bibinfo  {journal} {Phys.
			Plasmas}\ }\textbf {\bibinfo {volume} {25}},\ \bibinfo {pages} {022303}
	(\bibinfo {year} {2018})}\BibitemShut {NoStop}%
\bibitem [{\citenamefont {{Stawarz}}\ \emph {et~al.}(2019)\citenamefont
	{{Stawarz}}, \citenamefont {{Eastwood}}, \citenamefont {{Phan}},
	\citenamefont {{Gingell}}, \citenamefont {{Shay}}, \citenamefont {{Burch}},
	\citenamefont {{Ergun}}, \citenamefont {{Giles}}, \citenamefont {{Gershman}},
	\citenamefont {{Le Contel}}, \citenamefont {{Lindqvist}}, \citenamefont
	{{Russell}}, \citenamefont {{Strangeway}}, \citenamefont {{Torbert}},
	\citenamefont {{Argall}}, \citenamefont {{Fischer}}, \citenamefont
	{{Magnes}},\ and\ \citenamefont {{Franci}}}]{Stawarz2019ApJL}%
\BibitemOpen
\bibfield  {author} {\bibinfo {author} {\bibfnamefont {J.~E.}\ \bibnamefont
		{{Stawarz}}}, \bibinfo {author} {\bibfnamefont {J.~P.}\ \bibnamefont
		{{Eastwood}}}, \bibinfo {author} {\bibfnamefont {T.~D.}\ \bibnamefont
		{{Phan}}}, \bibinfo {author} {\bibfnamefont {I.~L.}\ \bibnamefont
		{{Gingell}}}, \bibinfo {author} {\bibfnamefont {M.~A.}\ \bibnamefont
		{{Shay}}}, \bibinfo {author} {\bibfnamefont {J.~L.}\ \bibnamefont {{Burch}}},
	\bibinfo {author} {\bibfnamefont {R.~E.}\ \bibnamefont {{Ergun}}}, \bibinfo
	{author} {\bibfnamefont {B.~L.}\ \bibnamefont {{Giles}}}, \bibinfo {author}
	{\bibfnamefont {D.~J.}\ \bibnamefont {{Gershman}}}, \bibinfo {author}
	{\bibfnamefont {O.}~\bibnamefont {{Le Contel}}}, \bibinfo {author}
	{\bibfnamefont {P.~A.}\ \bibnamefont {{Lindqvist}}}, \bibinfo {author}
	{\bibfnamefont {C.~T.}\ \bibnamefont {{Russell}}}, \bibinfo {author}
	{\bibfnamefont {R.~J.}\ \bibnamefont {{Strangeway}}}, \bibinfo {author}
	{\bibfnamefont {R.~B.}\ \bibnamefont {{Torbert}}}, \bibinfo {author}
	{\bibfnamefont {M.~R.}\ \bibnamefont {{Argall}}}, \bibinfo {author}
	{\bibfnamefont {D.}~\bibnamefont {{Fischer}}}, \bibinfo {author}
	{\bibfnamefont {W.}~\bibnamefont {{Magnes}}}, \ and\ \bibinfo {author}
	{\bibfnamefont {L.}~\bibnamefont {{Franci}}},\ }\href {\doibase
	10.3847/2041-8213/ab21c8} {\bibfield  {journal} {\bibinfo  {journal} {The
			Astrophysical Journal Letters}\ }\textbf {\bibinfo {volume} {877}},\ \bibinfo
	{eid} {L37} (\bibinfo {year} {2019})}\BibitemShut {NoStop}%
\bibitem [{\citenamefont {Akhavan-Tafti}\ \emph {et~al.}(2018)\citenamefont
	{Akhavan-Tafti}, \citenamefont {Slavin}, \citenamefont {Le}, \citenamefont
	{Eastwood}, \citenamefont {Strangeway}, \citenamefont {Russell},
	\citenamefont {Nakamura}, \citenamefont {Baumjohann}, \citenamefont
	{Torbert}, \citenamefont {Giles}, \citenamefont {Gershman},\ and\
	\citenamefont {Burch}}]{Akhavan-Tafti2018GRL}%
\BibitemOpen
\bibfield  {author} {\bibinfo {author} {\bibfnamefont {M.}~\bibnamefont
		{Akhavan-Tafti}}, \bibinfo {author} {\bibfnamefont {J.~A.}\ \bibnamefont
		{Slavin}}, \bibinfo {author} {\bibfnamefont {G.}~\bibnamefont {Le}}, \bibinfo
	{author} {\bibfnamefont {J.~P.}\ \bibnamefont {Eastwood}}, \bibinfo {author}
	{\bibfnamefont {R.~J.}\ \bibnamefont {Strangeway}}, \bibinfo {author}
	{\bibfnamefont {C.~T.}\ \bibnamefont {Russell}}, \bibinfo {author}
	{\bibfnamefont {R.}~\bibnamefont {Nakamura}}, \bibinfo {author}
	{\bibfnamefont {W.}~\bibnamefont {Baumjohann}}, \bibinfo {author}
	{\bibfnamefont {R.~B.}\ \bibnamefont {Torbert}}, \bibinfo {author}
	{\bibfnamefont {B.~L.}\ \bibnamefont {Giles}}, \bibinfo {author}
	{\bibfnamefont {D.~J.}\ \bibnamefont {Gershman}}, \ and\ \bibinfo {author}
	{\bibfnamefont {J.~L.}\ \bibnamefont {Burch}},\ }\href {\doibase
	10.1002/2017JA024681} {\bibfield  {journal} {\bibinfo  {journal} {Journal of
			Geophysical Research: Space Physics}\ }\textbf {\bibinfo {volume} {123}},\
	\bibinfo {pages} {1224} (\bibinfo {year} {2018})}\BibitemShut {NoStop}%
\bibitem [{\citenamefont {{Robert}}\ \emph {et~al.}(1998)\citenamefont
	{{Robert}}, \citenamefont {{Dunlop}}, \citenamefont {{Roux}},\ and\
	\citenamefont {{Chanteur}}}]{Robert1998ISSI}%
\BibitemOpen
\bibfield  {author} {\bibinfo {author} {\bibfnamefont {P.}~\bibnamefont
		{{Robert}}}, \bibinfo {author} {\bibfnamefont {M.~W.}\ \bibnamefont
		{{Dunlop}}}, \bibinfo {author} {\bibfnamefont {A.}~\bibnamefont {{Roux}}}, \
	and\ \bibinfo {author} {\bibfnamefont {G.}~\bibnamefont {{Chanteur}}},\
}\href@noop {} {\bibfield  {journal} {\bibinfo  {journal} {ISSI Scientific
		Reports Series}\ }\textbf {\bibinfo {volume} {1}},\ \bibinfo {pages} {395}
(\bibinfo {year} {1998})}\BibitemShut {NoStop}%
\bibitem [{\citenamefont {Bandyopadhyay}\ \emph {et~al.}(2018)\citenamefont
	{Bandyopadhyay}, \citenamefont {Chasapis}, \citenamefont {Chhiber},
	\citenamefont {Parashar}, \citenamefont {Matthaeus}, \citenamefont {Shay},
	\citenamefont {Maruca}, \citenamefont {Burch}, \citenamefont {Moore},
	\citenamefont {Pollock}, \citenamefont {Giles}, \citenamefont {Paterson},
	\citenamefont {Dorelli}, \citenamefont {Gershman}, \citenamefont {Torbert},
	\citenamefont {Russell},\ and\ \citenamefont
	{Strangeway}}]{Bandyopadhyay2018bApJ}%
\BibitemOpen
\bibfield  {author} {\bibinfo {author} {\bibfnamefont {R.}~\bibnamefont
		{Bandyopadhyay}}, \bibinfo {author} {\bibfnamefont {A.}~\bibnamefont
		{Chasapis}}, \bibinfo {author} {\bibfnamefont {R.}~\bibnamefont {Chhiber}},
	\bibinfo {author} {\bibfnamefont {T.~N.}\ \bibnamefont {Parashar}}, \bibinfo
	{author} {\bibfnamefont {W.~H.}\ \bibnamefont {Matthaeus}}, \bibinfo {author}
	{\bibfnamefont {M.~A.}\ \bibnamefont {Shay}}, \bibinfo {author}
	{\bibfnamefont {B.~A.}\ \bibnamefont {Maruca}}, \bibinfo {author}
	{\bibfnamefont {J.~L.}\ \bibnamefont {Burch}}, \bibinfo {author}
	{\bibfnamefont {T.~E.}\ \bibnamefont {Moore}}, \bibinfo {author}
	{\bibfnamefont {C.~J.}\ \bibnamefont {Pollock}}, \bibinfo {author}
	{\bibfnamefont {B.~L.}\ \bibnamefont {Giles}}, \bibinfo {author}
	{\bibfnamefont {W.~R.}\ \bibnamefont {Paterson}}, \bibinfo {author}
	{\bibfnamefont {J.}~\bibnamefont {Dorelli}}, \bibinfo {author} {\bibfnamefont
		{D.~J.}\ \bibnamefont {Gershman}}, \bibinfo {author} {\bibfnamefont {R.~B.}\
		\bibnamefont {Torbert}}, \bibinfo {author} {\bibfnamefont {C.~T.}\
		\bibnamefont {Russell}}, \ and\ \bibinfo {author} {\bibfnamefont {R.~J.}\
		\bibnamefont {Strangeway}},\ }\href {\doibase
	https://doi.org/10.3847/1538-4357/aade04} {\bibfield  {journal} {\bibinfo
		{journal} {The Astrophysical Journal}\ }\textbf {\bibinfo {volume} {866}},\
	\bibinfo {pages} {106} (\bibinfo {year} {2018})}\BibitemShut {NoStop}%
\bibitem [{\citenamefont {Jim\'enez}\ \emph {et~al.}(1993)\citenamefont
	{Jim\'enez}, \citenamefont {Wray}, \citenamefont {Saffman},\ and\
	\citenamefont {Rogallo}}]{jimenez1993JFM}%
\BibitemOpen
\bibfield  {author} {\bibinfo {author} {\bibfnamefont {J.}~\bibnamefont
		{Jim\'enez}}, \bibinfo {author} {\bibfnamefont {A.~A.}\ \bibnamefont {Wray}},
	\bibinfo {author} {\bibfnamefont {P.~G.}\ \bibnamefont {Saffman}}, \ and\
	\bibinfo {author} {\bibfnamefont {R.~S.}\ \bibnamefont {Rogallo}},\ }\href
{\doibase 10.1017/S0022112093002393} {\bibfield  {journal} {\bibinfo
		{journal} {Journal of Fluid Mechanics}\ }\textbf {\bibinfo {volume} {255}},\
	\bibinfo {pages} {65–90} (\bibinfo {year} {1993})}\BibitemShut {NoStop}%
\bibitem [{\citenamefont {Blackburn}\ \emph {et~al.}(1996)\citenamefont
	{Blackburn}, \citenamefont {Mansour},\ and\ \citenamefont
	{Cantwell}}]{Blackburn1996JFM}%
\BibitemOpen
\bibfield  {author} {\bibinfo {author} {\bibfnamefont {H.~M.}\ \bibnamefont
		{Blackburn}}, \bibinfo {author} {\bibfnamefont {N.~N.}\ \bibnamefont
		{Mansour}}, \ and\ \bibinfo {author} {\bibfnamefont {B.~J.}\ \bibnamefont
		{Cantwell}},\ }\href {\doibase 10.1017/S0022112096001802} {\bibfield
	{journal} {\bibinfo  {journal} {Journal of Fluid Mechanics}\ }\textbf
	{\bibinfo {volume} {310}},\ \bibinfo {pages} {269–292} (\bibinfo {year}
	{1996})}\BibitemShut {NoStop}%
\bibitem [{\citenamefont {Yang}\ \emph {et~al.}(2017)\citenamefont {Yang},
	\citenamefont {Matthaeus}, \citenamefont {Shi}, \citenamefont {Wan},\ and\
	\citenamefont {Chen}}]{Yang2017PoF}%
\BibitemOpen
\bibfield  {author} {\bibinfo {author} {\bibfnamefont {Y.}~\bibnamefont
		{Yang}}, \bibinfo {author} {\bibfnamefont {W.~H.}\ \bibnamefont {Matthaeus}},
	\bibinfo {author} {\bibfnamefont {Y.}~\bibnamefont {Shi}}, \bibinfo {author}
	{\bibfnamefont {M.}~\bibnamefont {Wan}}, \ and\ \bibinfo {author}
	{\bibfnamefont {S.}~\bibnamefont {Chen}},\ }\href {\doibase
	10.1063/1.4979068} {\bibfield  {journal} {\bibinfo  {journal} {Physics of
			Fluids}\ }\textbf {\bibinfo {volume} {29}},\ \bibinfo {pages} {035105}
	(\bibinfo {year} {2017})}\BibitemShut {NoStop}%
\bibitem [{\citenamefont {Chasapis}\ \emph
	{et~al.}(2018{\natexlab{b}})\citenamefont {Chasapis}, \citenamefont
	{Matthaeus}, \citenamefont {Parashar}, \citenamefont {Wan}, \citenamefont
	{Haggerty}, \citenamefont {Pollock}, \citenamefont {Giles}, \citenamefont
	{Paterson}, \citenamefont {Dorelli}, \citenamefont {Gershman}, \citenamefont
	{Torbert}, \citenamefont {Russell}, \citenamefont {Lindqvist}, \citenamefont
	{Khotyaintsev}, \citenamefont {Moore}, \citenamefont {Ergun},\ and\
	\citenamefont {Burch}}]{Chasapis2018ApJL}%
\BibitemOpen
\bibfield  {author} {\bibinfo {author} {\bibfnamefont {A.}~\bibnamefont
		{Chasapis}}, \bibinfo {author} {\bibfnamefont {W.~H.}\ \bibnamefont
		{Matthaeus}}, \bibinfo {author} {\bibfnamefont {T.~N.}\ \bibnamefont
		{Parashar}}, \bibinfo {author} {\bibfnamefont {M.}~\bibnamefont {Wan}},
	\bibinfo {author} {\bibfnamefont {C.~C.}\ \bibnamefont {Haggerty}}, \bibinfo
	{author} {\bibfnamefont {C.~J.}\ \bibnamefont {Pollock}}, \bibinfo {author}
	{\bibfnamefont {B.~L.}\ \bibnamefont {Giles}}, \bibinfo {author}
	{\bibfnamefont {W.~R.}\ \bibnamefont {Paterson}}, \bibinfo {author}
	{\bibfnamefont {J.}~\bibnamefont {Dorelli}}, \bibinfo {author} {\bibfnamefont
		{D.~J.}\ \bibnamefont {Gershman}}, \bibinfo {author} {\bibfnamefont {R.~B.}\
		\bibnamefont {Torbert}}, \bibinfo {author} {\bibfnamefont {C.~T.}\
		\bibnamefont {Russell}}, \bibinfo {author} {\bibfnamefont {P.~A.}\
		\bibnamefont {Lindqvist}}, \bibinfo {author} {\bibfnamefont {Y.}~\bibnamefont
		{Khotyaintsev}}, \bibinfo {author} {\bibfnamefont {T.~E.}\ \bibnamefont
		{Moore}}, \bibinfo {author} {\bibfnamefont {R.~E.}\ \bibnamefont {Ergun}}, \
	and\ \bibinfo {author} {\bibfnamefont {J.~L.}\ \bibnamefont {Burch}},\ }\href
{\doibase 10.3847/2041-8213/aaadf8} {\bibfield  {journal} {\bibinfo
		{journal} {The Astrophysical Journal Letters}\ }\textbf {\bibinfo {volume}
		{856}},\ \bibinfo {pages} {L19} (\bibinfo {year}
	{2018}{\natexlab{b}})}\BibitemShut {NoStop}%
\bibitem [{\citenamefont {Verma}(2004)}]{Verma04}%
\BibitemOpen
\bibfield  {author} {\bibinfo {author} {\bibfnamefont {M.~K.}\ \bibnamefont
		{Verma}},\ }\href {\doibase 10.1016/j.physrep.2004.07.007} {\bibfield
	{journal} {\bibinfo  {journal} {Physics Report}\ }\textbf {\bibinfo {volume}
		{401}},\ \bibinfo {pages} {229} (\bibinfo {year} {2004})}\BibitemShut
{NoStop}%
\bibitem [{\citenamefont {Coburn}\ \emph {et~al.}(2015)\citenamefont {Coburn},
	\citenamefont {Forman}, \citenamefont {Smith}, \citenamefont {Vasquez},\ and\
	\citenamefont {Stawarz}}]{Coburn2014PRS}%
\BibitemOpen
\bibfield  {author} {\bibinfo {author} {\bibfnamefont {J.~T.}\ \bibnamefont
		{Coburn}}, \bibinfo {author} {\bibfnamefont {M.~A.}\ \bibnamefont {Forman}},
	\bibinfo {author} {\bibfnamefont {C.~W.}\ \bibnamefont {Smith}}, \bibinfo
	{author} {\bibfnamefont {B.~J.}\ \bibnamefont {Vasquez}}, \ and\ \bibinfo
	{author} {\bibfnamefont {J.~E.}\ \bibnamefont {Stawarz}},\ }\href {\doibase
	10.1098/rsta.2014.0150} {\bibfield  {journal} {\bibinfo  {journal}
		{Philosophical Transactions Royal Society London A: Mathematical, Physical
			and Engineering Sciences}\ }\textbf {\bibinfo {volume} {373}} (\bibinfo
	{year} {2015}),\ 10.1098/rsta.2014.0150}\BibitemShut {NoStop}%
\end{thebibliography}



\end{document}